\documentclass[galaxies,article,accept,moreauthors,pdftex]{Definitions/mdpi} 
\usepackage{lscape}
\usepackage{longtable}
\usepackage{amsmath,siunitx}
\usepackage{color}

\graphicspath{{fig/}}
\setitemize{parsep=6pt,itemsep=0pt,leftmargin=*,labelsep=5.5mm}
\setenumerate{parsep=6pt,itemsep=0pt,leftmargin=*,labelsep=5.5mm}
\setlist[description]{itemsep=0mm}   

\firstpage{1} 
\pubvolume{8}
\issuenum{1}
\articlenumber{16}
\pubyear{2020}
\copyrightyear{2020}
\history{Received: 22 November 2019; Accepted: 6 February 2020; Published: 17 February 2020}
\updates{yes} 

\Title{Frequency of Planets in Binaries }

\Author{{Mariangela Bonavita} $^{1,2,}$*\orcidA and {Silvano Desidera} $^{3}$\orcidB}
\AuthorNames{Mariangela Bonavita, Silvano Desidera}

\address{%
$^{1}$ \quad SUPA, Institute for Astronomy, University of Edinburgh, Blackford Hill, Edinburgh EH9 3HJ, UK\\
$^{2}$ \quad Centre for Exoplanet Science, University of Edinburgh, Edinburgh EH9 3HJ, UK:\\
$^{3}$ \quad INAF---Osservatorio Astronomico di Padova, Vicolo dell'Osservatorio 5, I-35122, Padova, Italy; silvano.desidera@inaf.it}

\corres{Correspondence: mbonav@roe.ac.uk}

\abstract{The frequency of planets in binaries is an important issue in the field of extrasolar planet studies because of its relevance in the estimation of the global planet population of our galaxy and the clues it can give to our understanding of planet formation and evolution.
Multiple stars have often been excluded from exoplanet searches, especially those performed using the radial velocity technique, due to the technical challenges posed by such targets. 
As a consequence and despite recent efforts, our knowledge of the frequency of planets in multiple stellar systems is still rather incomplete. 
On the other hand, the lack of knowledge about the binarity at the time of the compilation of the target samples means that our estimate of the planet frequency around single stars could be tainted by the presence of unknown binaries, especially if these objects have a different behavior in terms of planet occurrence. 
In a previous work we investigated the binarity of the objects included in the Uniform Detectability sample defined by Fisher and Valenti (2005), showing how more than 20\% of their targets were, in fact, not single stars.  
Here, we present an update of this census, made possible mainly by the information now available thanks to the second Gaia Data Release. 
The new binary sample includes a total of 313 systems, of which 114 were added through this work. We were also able to significantly improve the estimates of masses and orbital parameters for most of the pairs in the original list, especially those at close separations. A few new systems with
white dwarf companions were also identified. The results of the new analysis are in good agreement with the findings of our previous work, confirming the lack of difference in the overall planet frequency between binaries and single stars but suggesting a decrease in the planet frequency for very close pairs.}


\begin{document}

\section{Introduction}

More than 70\% of massive early-type stars \citep{Kouwenhoven2007, Peter2012} and 50\%--60\% of solar-type stars \citep{DM91, raghavan2010} are observed in binary or higher order multiple systems, with the fraction decreasing to 30\%--40\% for M-stars \citep{Fischer1992, Delfosse2004, Janson2012}. Therefore, in order to properly assess the global frequency of planets, it is crucial to consider the role of stellar companions.
Any difference between frequencies or properties of planets in single and multiple systems would shed light on the effects of the presence of the companion star on the planet formation mechanisms, particularly in the case of very close binaries
\cite{marzari2019}. This review studies the impact of binarity on the planet formation process.

On the observational side, several serendipitous discoveries and a few recent dedicated surveys have revealed a significant number of exoplanets both around individual components of binary systems (circumstellar or S-like configurations; see, e.g., \citep[][]{Eggenberger2007, Mugrauer2007, DesideraBarbieri2007}) and around both components of tight binaries (circumbinary or P-like configurations; see, e.g., \citep[][]{doyle2011,orosz2012,beuermann2010}).
These discoveries have triggered several studies investigating the impact of stellar binarity on planet formation and on the planet demographics; see, e.g., \citep[][]{DesideraBarbieri2007,Eggenberger2007,Eggenberger2011,Daemgen2009,Adams2012,Adams2013,Ginski2012}). Some of these studies highlighted a strong deficit of binary companions within $\sim$50--100 au for planet hosts \citep{Roell2012,Bergfors2013,Wang2014a,Wang2014b,kraus2016}, while others seem to suggest a low impact of the presence of stellar companions on planet formation \citep{horch2014}.
A first work on planet candidates from the Transiting Exoplanets Survey Satellite (TESS) indicates a paucity of binaries with separations of closer than 100 au and an overabundance of binaries around stars with Hot Jupiter candidates \citep{ziegler2019}, confirming the recent findings by \citep{fontanive2019}, who pointed out how giant planets and brown dwarf desert inhabitants are almost exclusively observed in multiple systems.
Overall, while it is clear that both the properties and the occurrence of planets in binaries can be different with respect to those orbiting single stars, the precise nature and extent of such differences remain unclear.
A robust statistical analysis, based on well-defined samples with complete or well-characterized detection limits both for planetary companions and for stellar companions over the full separation range, is therefore crucial to obtaining useful constraints to this problem. 

The work by \cite{BD07} (hereafter BD07) represents one of the first attempts in this direction. They~performed a search for stellar companions of the stars in the Uniform Detectability (UD) sample by~\cite{FV05} (hereafter FV05). The UD sample includes a total of 850 stars from the Keck, Lick, and Anglo Australian Telescope (AAT) radial velocity (RV) surveys for which the detectability of planets is considered complete for orbital periods shorter than four years and RV semi-amplitudes larger than 30~m/s. A total of 50 stars in the sample have at least one planet within these limits.
It should be noticed that these surveys have biases against binaries, excluding visual binaries with projected separations of smaller than 2$^{\prime\prime}$ and spectroscopic binaries known at the time of the sample selection. 
Nevertheless, thanks to the completeness of planet detection and the large sample size, the~UD sample can still be used to draw conclusions about the frequency of planets in binary stars.
Via a thorough search of both the literature and data available at that time, BD07 highlighted the presence of 199 multiple systems in the UD sample, spanning a wide range of separations. 
The~resulting statistical analysis pointed towards a lower frequency of planets among these binaries with respect to single stars, but only for those with separations of below 100~au.

The UD sample is still a reference sample for statistics of giant planets, as no other similar works have been published from other surveys. 
The accumulation of data in the last years and, in particular, the Gaia DR2 release \cite{gaia_dr2} now allow for a significant improvement of the census of binaries both at wide and short separations. 
Minor updates to the census of planets within the UD boundaries should also be applied following the recent literature.

The purpose of this study is therefore to present a much-needed update of the work by BD07.

The paper is organized as follows: The methods used to update the UD binary sub-sample are presented in Section~\ref{sed:udupdates}, while Section~\ref{sec:results} summarizes the results and their implications on the frequency of planetary companions. Finally, the conclusion of the study is presented in Section~\ref{sec:conclusions}.

\section{Updates of the UD Binary Sample}
\label{sed:udupdates}
The completeness, in terms of binary detection, around FGK stars, such those included in the UD sample, has significantly improved since the publication of the census performed by BD07. 
The~FV05 UD sample is still used as reference for the statistics of planets around FGK stars from the RV technique but, besides a minor update presented in \cite{bonavita2010}, there has been no further work aimed at characterizing the binary frequency among its targets, despite all of the new information available. 
We therefore decided to repeat the search for binaries among the UD stars in order to re-assess the frequency of planets around single stars and stars in multiple systems at various separations, thus updating the work presented by BD07. 

Apart from the extension of the sample, the new search was also necessary for clarifying the nature of several tens of targets which were included in the UD binary sample by BD07 only because of the presence of long-term RV trends (mostly from \cite{Nidever2002}) or astrometric signatures of binarity (proper motion difference between Tycho and Hipparcos, \cite{MakarovKaplan2005}), without many details on the masses and separation of the unseen companions.
For most cases, these ambiguities are now solved, thanks to the determination of the RV orbital solution and/or the direct detection of the companions.

In addition to several studies published about individual systems, the new census of UD binaries was possible mainly thanks to the availability of data from: 

\begin{itemize}
    \item Gaia DR2 \cite{gaia_dr2}, which provided a major source of directly detected companions down to moderately small separations and faint magnitudes \citep{brandeker2019}. It also allowed for the determination of physical associations from parallaxes and proper motions of the components, and for the identification of additional
    $\Delta \mu$ targets (see below).
    \item The works by \cite{Patel2007,Jenkins2015}, which included spectroscopic orbital solutions for binaries detected in the surveys that formed the UD sample. 
    \item Several dedicated adaptive optics (AO) surveys targeting stars with RV planets \citep{Mugrauer2007,Eggenberger2007,moutou2017}, objects with RV long-term trends \citep{Kane2019,crepp2014}, and stars with astrometric acceleration from Tycho and Hipparcos \citep{tokovinin2012}, all published after BD07. 
    \item The full RV time series for the Keck and Lick planet search surveys which were published in \cite{Butler2017} and \cite{Fischer2014}. These allowed us to assess whether the astrometric trends observed for some of the UD stars were due to unknown massive companions and, if so, to confirm their stellar or planetary natures. We were also able to to derive preliminary orbital solutions for a few binaries not included in previous works\footnote{\cite{tal-or2019}{ identified low-amplitude (about 1 m/s)} systematic effects in the Keck-HIRES RV time series from \cite{Butler2017}
    and published the time series corrected for these effects. The impact of these systematic errors is negligible for the purpose of our work given their low amplitude. In~any case, we adopted the corrected time series
    for our orbit-fitting and trend evaluations.}.
\end{itemize}

The impact of each of these sources on the final updated UD binary sub-sample is described in detail in the following sections. 

\subsection{New Spectroscopic Orbital Solutions}

The availability of RV time series from the Keck and Lick RV surveys allowed us to investigate the case of five previously unpublished spectroscopic binaries.
The orbital parameters, reported in Table~\ref{tab:newspec}, were derived according to the approach described in \cite{Desidera2011}. In most cases, the orbital periods are longer than the time baselines of the observations, making the retrieved orbital parameters preliminary but still useful in the context of our work.   
All of the companions are low-mass stars (minimum mass 0.1--0.25~$M_{\odot}$) with semi-major axes in the range of 6 to 16 au. 
\begin{table}[H]
\caption{\footnotesize Preliminary spectroscopic orbital solutions from the individual radial velocity time series published by~\cite{Fischer2014} for Lick and by \cite{Butler2017} for Keck surveys.}
\label{tab:newspec}
\begin{center}
\begin{tabular}{lcccccccc}
\toprule
\multirow{2}{*}{\textbf{HD}}       & \textbf{P}    & \textbf{K}      & \multirow{2}{*}{\textbf{ecc}} & \boldmath{$\omega$}  &  \textbf{T0} & \textbf{msini}    & \textbf{a}    & \multirow{2}{*}{\textbf{RV Source}} \\
         &\textbf{(d) }  & \textbf{(m/s)}  &      & \textbf{deg} & \textbf{JD-2450000}  & \boldmath{$M_{\odot}$}  & \textbf{(au)} &      \\
\midrule
HD 30649  &  23250 & 1752    & 0.56 &  310.2 & 3966.3  & 0.21 & 16.5 & Keck \\ 
HD 103829 &   4880 & 1439    & 0.27 &  360.0 &  1545.7 & 0.13 &  6.20 & Keck \\
HD 190771 &   8527 & 1139    & 0.53 &  103.6  & 3707.2 & 0.10 &  8.61 & Lick \\ 
HD 190387 &   5155 & 2081    & 0.53 &   74.6  & 4155.8 & 0.19 &  6.70 & Keck \\
HD 218101 &   5156 & 2684    & 0.58 &  161.3  & 5169.1 & 0.23 &  6.67 & Lick \\ 

\bottomrule
\end{tabular}
\end{center} 
\end{table}

\subsection{New Visual Binaries from Gaia}
\label{sec:gaiavis}
\textls[-15]{We used the method described in \cite{fontanive2019} to search for additional companions of the UD stars in the Gaia DR2 catalogue. For each star in the sample, we used a search radius corresponding to $10^4$~au, then selected the sources with relative parallaxes and at least one of the two proper motion components differing by no more than 20\%, with the second component being within 50\%. 
Using these criteria, we retrieved 103 pairs already included in the original BD07 compilation, either because of previous detection of the companion or due to the presence of a dynamic signature, and added 95~new~companions. }

As discussed in detail in \cite{fontanive2019}, these criteria are rather conservative, as they are meant to ensure that no spurious objects are included. This lead to 25 cases\footnote{Note that systems for which both components were included in the UD sample were counted twice.} in which companions already confirmed and included in BD07 were not retrieved in our cross-match with the Gaia catalogue, mainly because the proper motion or parallax difference between the components was larger than the adopted threshold.
These cases were checked individually and a suitable explanation for the discrepancies was found in nearly all of them. In most cases, either the pair is close enough to have relevant orbital motion, or~one of the components is itself a known close binary (14 and 10 stars, respectively). The remaining ambiguous case (HD 50281) is discussed in the Appendix~\ref{app:remarks}.
Cases like those discussed above should be relatively rare, and we do not expect them to lead to significant incompleteness in our census of binaries. In fact, very close companions, which would cause large proper motion differences, would have  already been included, thanks to other indicators such as long-term RV trends, and then checked individually. Wider companions are, in principle, more challenging to disentangle, but, as shown in Section~\ref{sec:results}, our final sample is nearly complete at separations larger than 2$^{\prime\prime}$.

\subsection{Dynamically Inferred Systems}
\label{sec:newdmu}
A significant discrepancy in proper motion measurements between catalogs of different time spans ($\Delta\mu$) is considered to be a good indication of the presence of a perturbing body, and has been successfully used in the past to select potential stellar binaries (e.g., Makarov and Kaplan 2005; Tokovinin et al. 2013).
A total of 38 objects were included in the original UD binary sample from BD07 solely because of the presence of significant $\Delta\mu$ between Tycho-II \citep{Tycho2} and Hipparcos \citep{1997hity.book.....P} or because of other dynamical signatures (RV trends, astrometric acceleration detected in Hipparcos, etc.). 
Thanks to the search described in Section~\ref{sec:gaiavis} or through the new information available in the literature, we~were able to confirm the binary nature of 27 of these. Such a high binary fraction among $\Delta\mu$ targets, in agreement with the recent results by \cite{fontanive2019b}, and the availability of new and more precise proper motion measurements justified a new search for UD stars with discrepancies in proper motion between Gaia DR2 \citep{gaia_dr2} and both Gaia DR1 TGAS \citep{GaiaDR1} and Tycho-II \citep{Tycho2} catalogs.
The objects for which the astrometric signature was compatible with the presence of planetary mass companions, including six of the BD07 $\Delta\mu$ pairs, were removed, leaving us with 65 new $\Delta\mu$ potential pairs.
Of these, 37 had already been included in the binary sub-sample based on the search described in Section~\ref{sec:gaiavis} or because of previously published results. 
We also systematically investigated the $\Delta \mu$ systems by evaluating the presence of confirmed RV planets or the presence of long-term RV trends, exploiting the published RV time series \citep{Fischer2014,Butler2017}.

\begin{figure}[H]
    \centering
    \includegraphics[width=7.7cm]{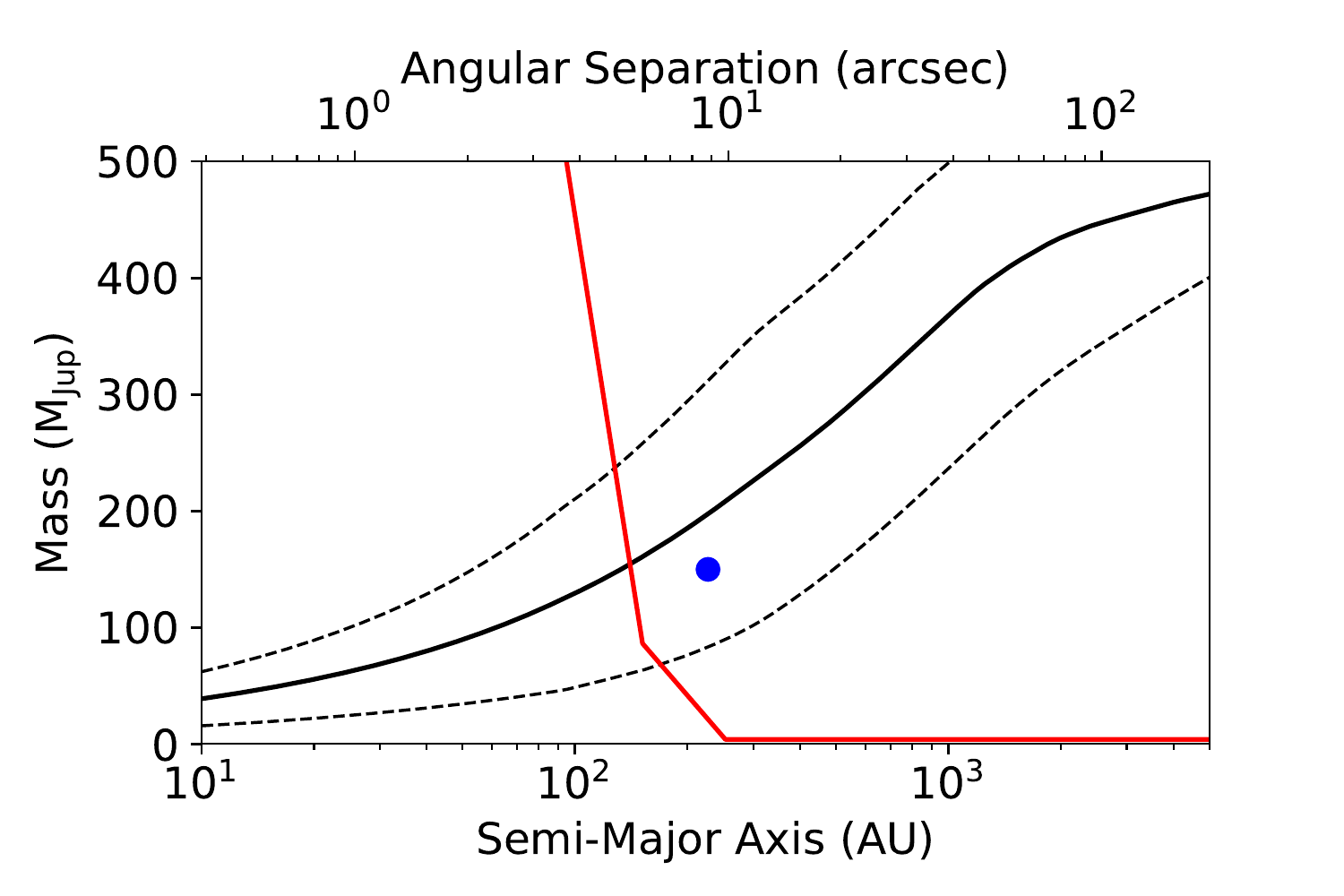}
    \includegraphics[width=7.7cm]{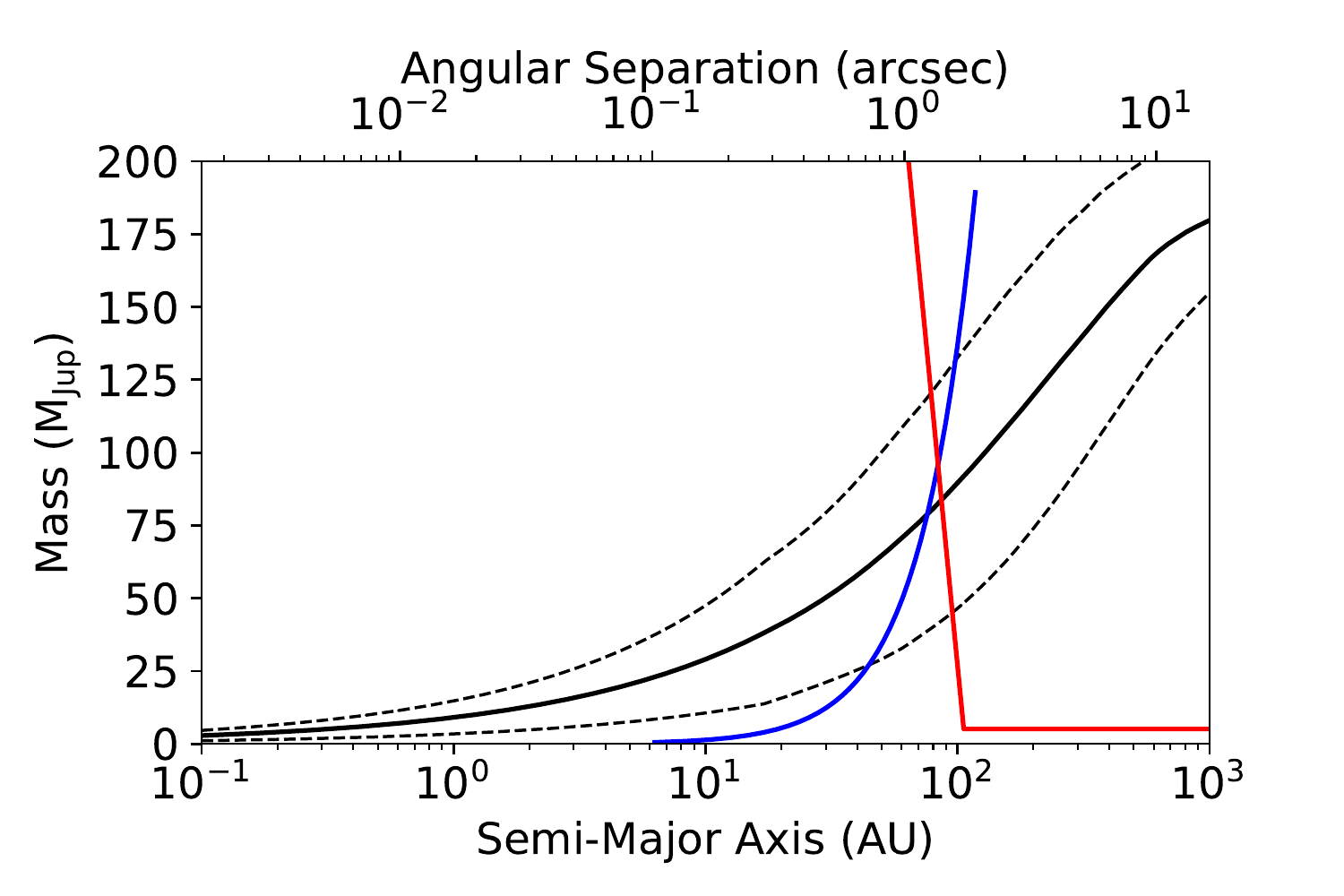}
    \caption{Mass vs. semi-major axes of companions compatible with the observed $\Delta\mu$ for HD~67458 (left panel) and HD~80913 (right panel), calculated using the COPAINS (Code for Orbital Characterization of Astrometrically Inferred New Systems; see \citep[][]{fontanive2019b} for details) and assuming a uniform distribution for the eccentricity. The red curve shows Gaia's sensitivity limits~\citep{fontanive2019, Ziegler2018}. The blue dot in the left panel shows the position of the companion to HD~67458 retrieved in Gaia DR2. The blue solid line in the right panel shows the position of the companions compatible with the RV trend observed by \cite{Kane2019} for HD~80913.}
    \label{fig:dmuplots}
\end{figure}

For the remaining objects with no confirmed companions, including the remaining six unconfirmed $\Delta\mu$ pairs from BD07, we used COPAINS (Code for Orbital Characterization of Astrometrically Inferred New Systems; see \citep[][]{fontanive2019b} for details) to estimate the mass and separation of a possible unseen companion compatible with the observed $\Delta\mu$. 
For consistency with what was later assumed for the evaluation of the critical semi-major axis for dynamical stability \citep{HolmanWiegert1999}, we adopted a uniform distribution for the eccentricity of the secondary (also in agreement with the results by \citep[][]{raghavan2010, tokovinin2012})\footnote{The validity of this choice is also further justified by the assumption that tidal circularization is not expected to play a significant role for binaries, such those in our sample. Close binaries with tidally circularized orbits are likely to have been excluded from the original survey samples because they are easier to discover, thanks to the short orbital periods and large RV amplitudes. Furthermore, active stars (as tidally-locked binaries) were typically excluded from the survey to avoid the negative effect of activity jitter of planet detectability.}.
The~left panel of Figure~\ref{fig:dmuplots} shows an example of the output of the code (black curves) as well as the expected detection limits for Gaia (red curve), calculated using the results of~\cite{Ziegler2018} and following the approach of \cite{fontanive2019}. 
The position of the companion identified in Gaia DR2 for the same target is marked by a blue dot. As~expected, the~secondary is both within the parameter space identified by COPAINS and also above Gaia's detection~limit. 

In case an RV trend was also detected for those targets, we used this information to further constrain the mass and separation of a companion to those compatible with both signatures. The right panel of Figure~\ref{fig:dmuplots} shows an example of the output of COPAINS for one of these objects, together with the values compatible with the RV trend calculated using the method described by \cite{torres1999} (blue curve). 
\vspace{-12pt}

\subsection{Revision of Individual Masses}
The availability of high-quality Gaia G-band magnitudes allowed for a systematic revision of stellar masses of the companions from mass--luminosity relationships.
To this end, we used the the tables by \cite{pecaut2013}\footnote{Version 2019.03.22 available at \url{http://www.pas.rochester.edu/~emamajek/EEM_dwarf_UBVIJHK_colors_Teff.txt}.} to perform a polynomial fit of the stellar mass and absolute G, J, and K magnitudes.
The~adopted solutions, displayed in Equations (\ref{eq:massG}) to (\ref{eq:massK}), were derived for stars of spectral types from F0 to L2 (intended as nominal substellar limits) and the values of the masses obtained have an estimated error of $\sim 0.01 M_{\odot}$. The actual uncertainty of the mass is most likely much higher and can vary according to, among other things, the distance of the target from the main sequence (both at young and old ages) and the stellar metallicity (not included in the tables by \cite{pecaut2013}). Although a proper derivation of the total uncertainties is challenging, we do not expect it to be higher than 0.05~$M_{\odot}$ for unevolved late-type stars, as the companions to the UD stars considered in this study.
\begin{align}
    Mass (G)=2.53615-0.549060*M_{G}+0.143017*M_{G}^2-0.0542299*M_{G}^3 +0.0127813*M_{G}^4 \notag\\
    -0.00163578*M_{G}^5 + 0.000113799*M_{G}^6 -4.06116e-06*M_{G}^7+5.82965e-08*M_{G}^8 \label{eq:massG} \\
    \mspace{150mu}
    \notag\\
    Mass (J)=3.95501-2.87761*M_{J}+1.58846*M_{J}^2-0.568685*M_{J}^3 + 0.125674*M_{J}^4 \notag \\
    -0.0169899*M_{J}^5+0.00136157*M_{J}^6 -5.91812e-05*M_{J}^7+1.07257e-06*M_{J}^8 \label{eq:massJ}\\
    \mspace{150mu}
    \notag\\    
    Mass (K)=6.17137-7.08845*M_{K}+4.79712*M_{K}^2-1.89004*M_{K}^3+0.446451*M_{K}^4 \notag \\
    -0.0641778*M_{K}^5+0.00549026*M_{K}^6-0.000256553*M_{K}^7+5.03656e-06*M_{K}^8 \label{eq:massK}
\end{align}
    
In the case of wide companions, the adopted mass value included in Table~\ref{tab:binary_UD} was calculated as the mean of the ones obtained from each of the solutions. The value obtained from Equation~(\ref{eq:massG}) was preferred for binaries with separations smaller than 7--10$^{\prime\prime}$, as the quality of 2MASS photometry degrades for close visual binaries. 
When detailed information on the system was available in the literature, the published values were preferred (these objects have $M_{flag}~=~c$ in Table~\ref{tab:binary_UD}, with the corresponding reference listed in the notes at the bottom). 

\subsection{Revision of the UD Sample}
\label{sec:udrevision}

We revised the classification of several stars with/without planets within the UD boundaries on the basis of the updated results from the RV time series or the determination of astrometric masses that moved a detected companion outside the planetary regime.
In more detail, we changed the status for the following targets with respect to BD07:
\begin{itemize}
    \item HD 196885 is put back among the stars with planets, as the RV planet is confirmed \citep{Correia2008, fischer2009}.
    \item HD 136118 was originally flagged as a star with planets, but the astrometric analysis by \cite{martioli2010} shows that the true mass of the companion (41 $M_{Jup}$) is significantly larger than the minimum mass from RV only and larger than our adopted limit for planetary mass companions. We therefore included this object among the stars without planets. 
    \item HD 137510 has a minimum mass above 24 $M_{Jup}$ \citep{diaz2012}, and is therefore classified as a star without planets.
    \item HD 159868 was considered as without planets in FV05. After the revision of the planetary orbits, a second planet was discovered \citep{luhn2019}. As one the planets fulfills the UD definition, the star is now considered to be with planets. It is not known to be a binary. 
\end{itemize}

\section{Results}
\label{sec:results}

\subsection{The Updated UD Binary Sample}

\begin{figure}[H]
    \centering
    \includegraphics[height=6cm]{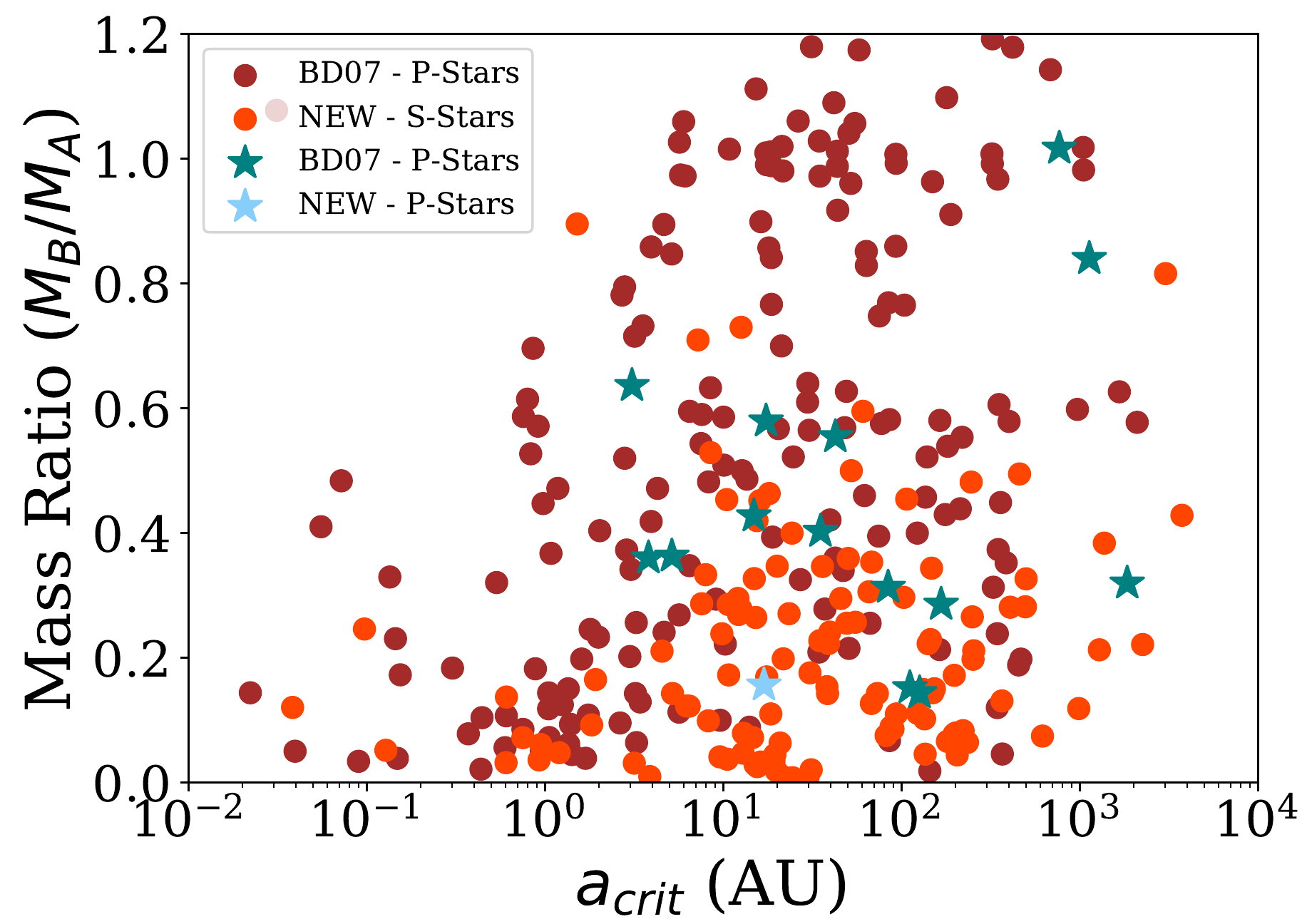}
    \caption{Critical semi-major axis vs. mass ratio ($M_A/M_B$) for the pairs in the Uniform Detectability (UD) sample with (stars) and without (filled circles) planetary companions. The new pairs not in BD07 are shown in light blue and orange, respectively.}
    \label{fig:qvsacrit}
\end{figure}

The full list of UD binaries as assembled above includes 313 of of the 850 stars in the sample. A~total of 114~targets were added with respect to BD07 and three were removed (physical association rejected by Gaia, possibility of planetary companion explaining the astrometric signature). 
For all of the pairs, we~derived the critical semi-major axis for dynamical stability (hereafter $a_{crit}$; see \cite{HolmanWiegert1999} and BD07 for details). When available, we exploited the binary orbits. For the cases in which only the projected separation was available, we used the same approach as \cite{bonavita2016} and estimated $a$(au) as $~\rho$($^{\prime\prime}$)$d$(pc), thus assuming a flat eccentricity distribution (see Section~\ref{sec:newdmu} for details).
In agreement with the assumption used for the semi-major axis calculation, an eccentricity value of 0.5 was adopted for the systems for which no information on the orbit was available\footnote{Note that this approach is slightly different from that adopted in BD07 and, therefore, the critical semi-major axis values cannot be compared directly.}. 
The properties of all of the objects in the updated UD binary sample are listed in Table \ref{tab:binary_UD}. Figure~\ref{fig:qvsacrit} shows the values of $a_{crit}$ as a function of the mass ratio ($M_A/M_B$) for all pairs, highlighting the newly added objects. 

\begin{figure}[H]
    \centering
    \includegraphics[height=6cm]{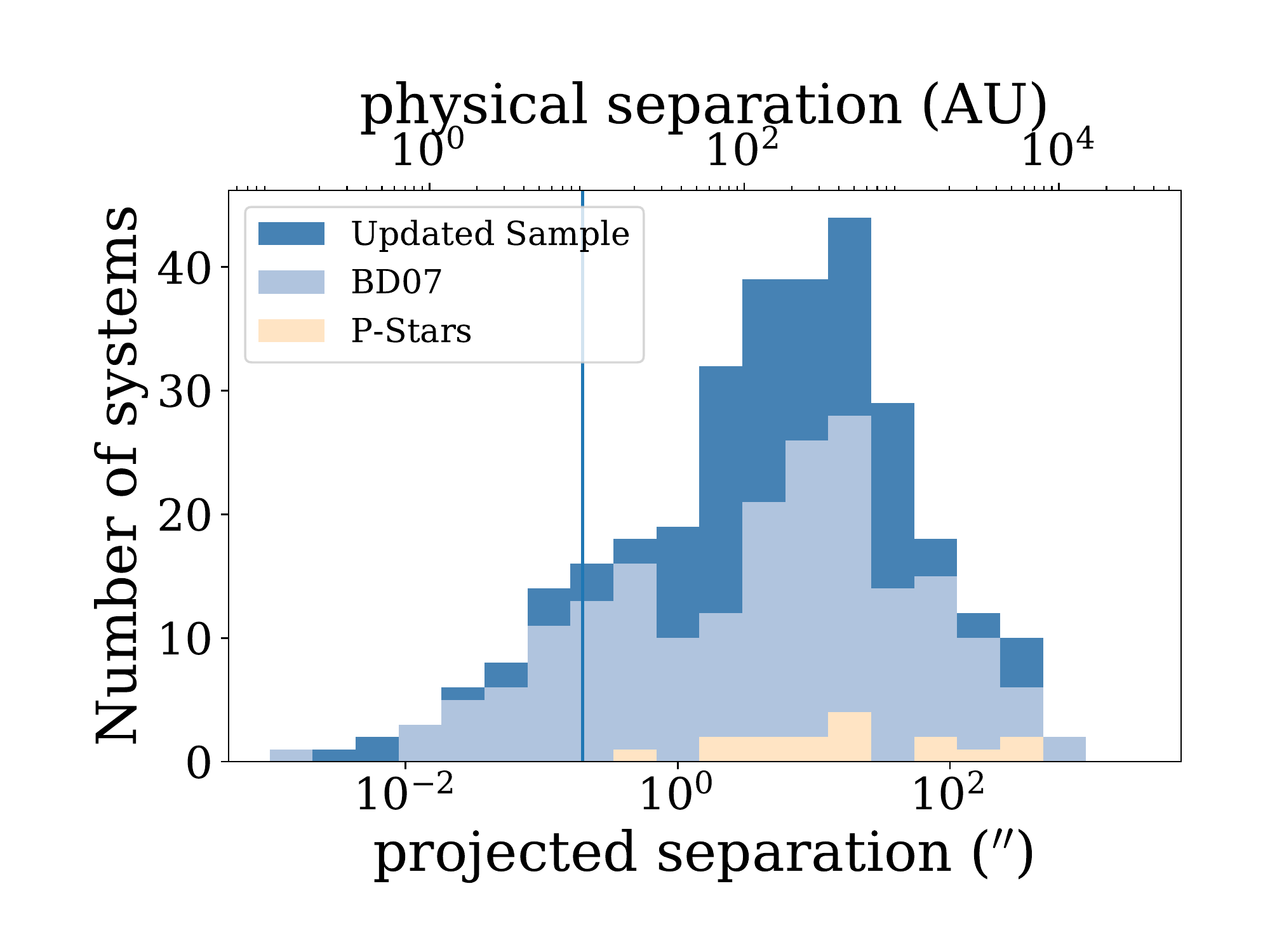}
    \caption{Histogram of the projected separation for the pairs in the updated (dark blue) and the original (light blue) UD binary sample. The number of stars with planets (P-Stars) in the final sample is shown in light orange. The solid line corresponds to $\rho = 2 ^{\prime\prime}$.}
    \label{fig:rhohist}
\end{figure}

Despite the increased size of the binary sub-sample, we still observe a lower overall frequency of binaries ($\sim37\%$ as opposed to $47\%$) compared to what was predicted by \cite{DM91} for a volume-limited sample of the same size as the UD. 
As discussed in BD07, this difference is most likely due to the fact that the input lists used to build the UD sample had an explicit bias against close binaries. 
According to~\cite{DM91}, in~a sample of the same size as the UD, we would have expected a total of 484 binaries, of which $\sim 45.7\%$ would have had projected separations below $2^{\prime\prime}$ and would therefore have been excluded from the UD sample. 
If restricted to binaries with $\rho > 2^{\prime\prime}$, our updated UD binary sample includes 211 pairs, thus implying a level of completeness of $\sim95\%$: A significant improvement with respect to the $62\%$ achieved in BD07, as also highlighted in Figure~\ref{fig:rhohist}.
\unskip

\begin{figure}[H]
    \centering
    \includegraphics[height=5.75cm]{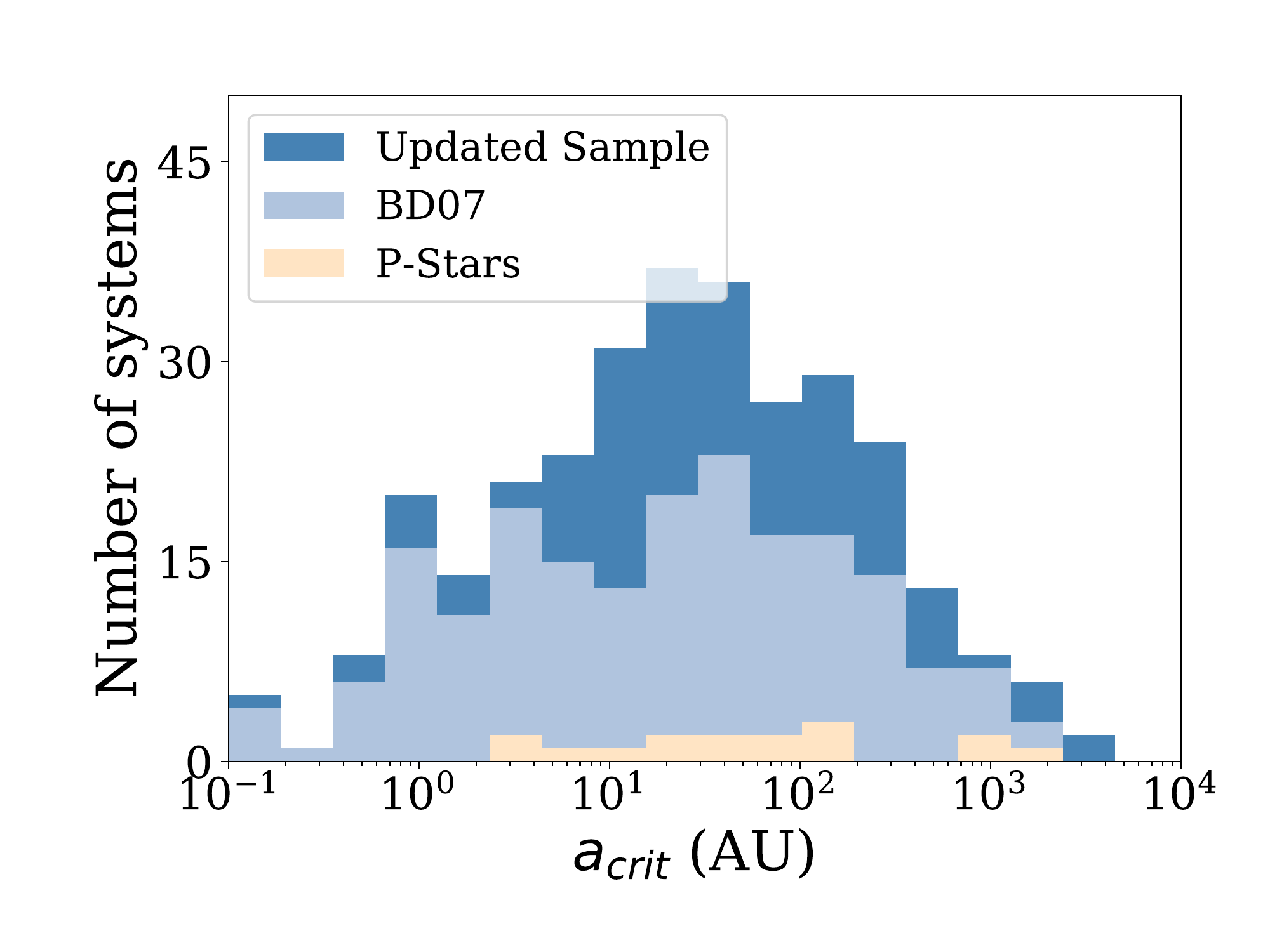}
    \includegraphics[height=5.75cm]{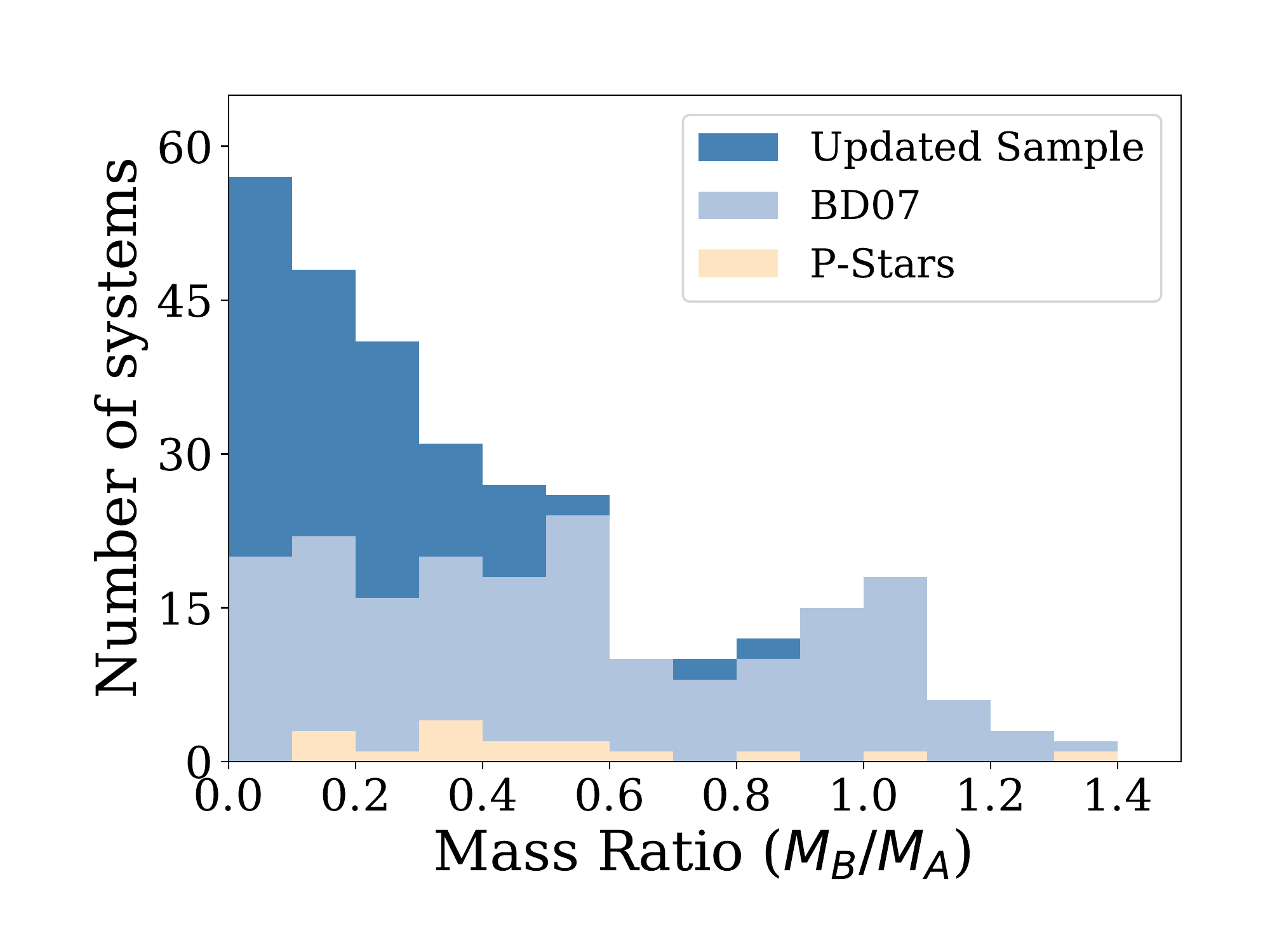}
    \caption{Histogram of the critical semi-major axis ($a_{crit}$, \textbf{left panel}) and mass ratio ($M_B/M_A$,  \textbf{right panel}) for the pairs in the updated (dark blue) and the original (light blue) UD binary sample. The~number of stars with planets (P-Stars) in the final sample is shown in light orange.}
    \label{fig:samplehist}
\end{figure}

 On the other hand, the fraction of close binaries did not significantly increase with respect to the original BD07 sample ($32.25\%$ as opposed to $30\%$). While it is true that a clear assessment of the incompleteness of the sample at such separations is hard to complete, we should note that the quality of the information available for these systems in the updated sample is much higher. 
Several systems in the original sample had been selected solely on the basis of dynamical signatures and were included in the lowest mass ratio and $a_{crit}$ bins due to the lack of information on the companion masses and orbits.
As~discussed in Section~\ref{sec:newdmu}, most of those systems were confirmed and can now be correctly placed in the appropriate bins. The fact that most of these systems had low-mass companions, undetectable at the time of the BD07 compilation, explains the clear increase of low-mass ratio systems shown in the right panel of Figure~\ref{fig:samplehist}.\vspace{-12pt}
 
\subsection{Frequency vs. Critical Semi-Major Axis for Dynamical Stability}
\label{sec:frequency}

With 16 of the 313 pairs in the updated UD binary sample hosting a planetary companion, we~estimated the global frequency of planets in the UD binary sample to be 5.1\%. This value is compatible, within the errors, with the the frequency of planets in the UD single-star sub-sample, which is 6.3\% (see Table~\ref{tab:freq}).

\textls[-15]{In order to investigate the dependence of the planet frequency on dynamical effect due to the presence of stellar companions, we divided the binary sample into some sub-samples according to the value of $a_{crit}$. This feature, in fact, includes both the orbital parameters and the mass ratio, representing the maximum value of the semi-major axis for stable planetary orbits around planet hosts (see \citep[][]{DM91} for~details).}

Table~\ref{tab:freq} shows the values of the frequency for different intervals of $a_{crit}$, as well as the values of the frequency in the full UD binary sample compared to those of the UD singles. 
Following the approach adopted in \cite{BD07}, the errors on the frequency values were estimated as: $\sigma_f= \left(N_{planets}^{-1/2} + N_{star}^{-1/2}\right)*\left(\frac{N_{planets}}{N_{stars}}\right)$.

\begin{figure}[H]
    \centering
    \includegraphics[width=7.75cm]{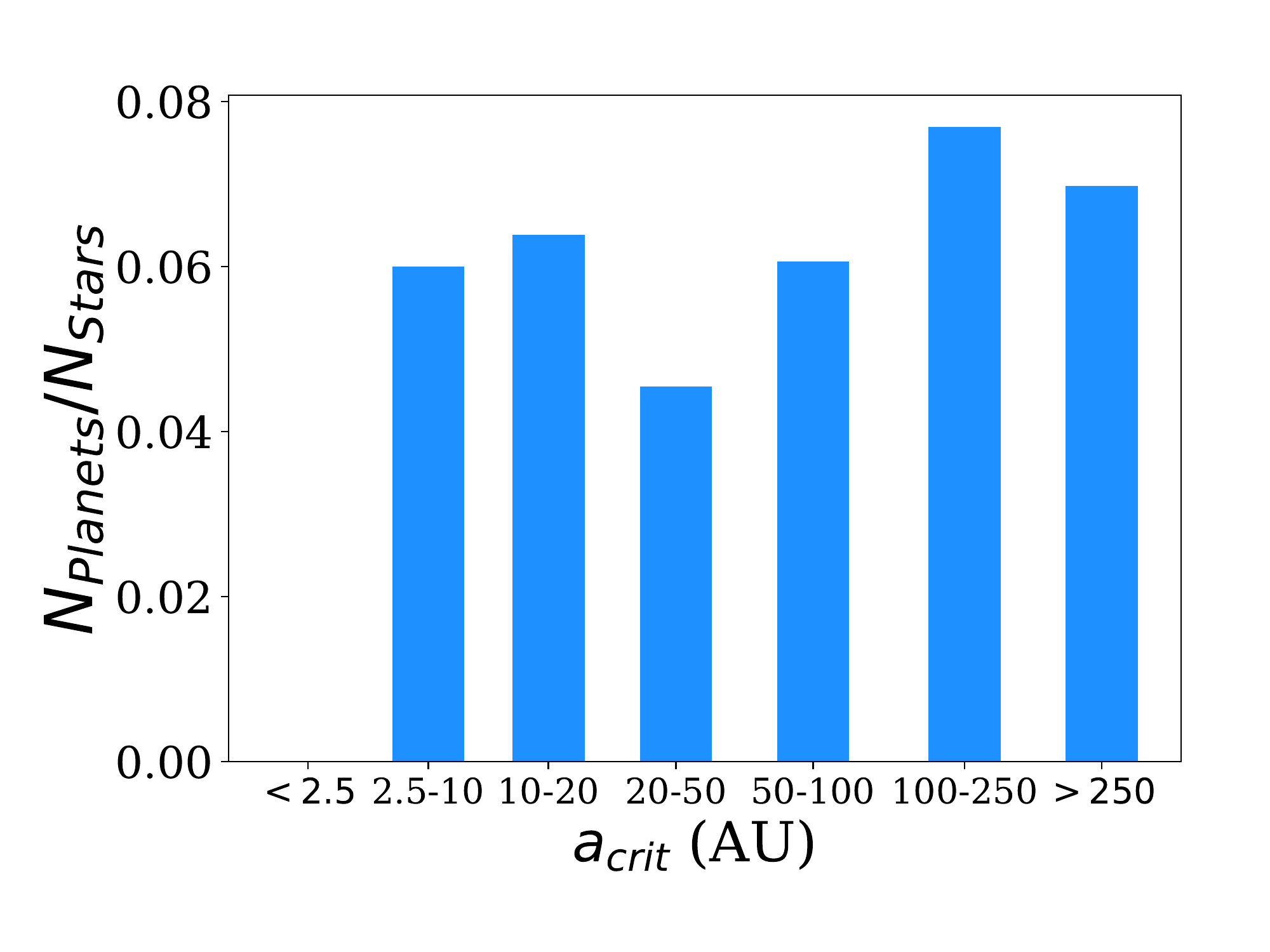}
    \includegraphics[width=7.75cm]{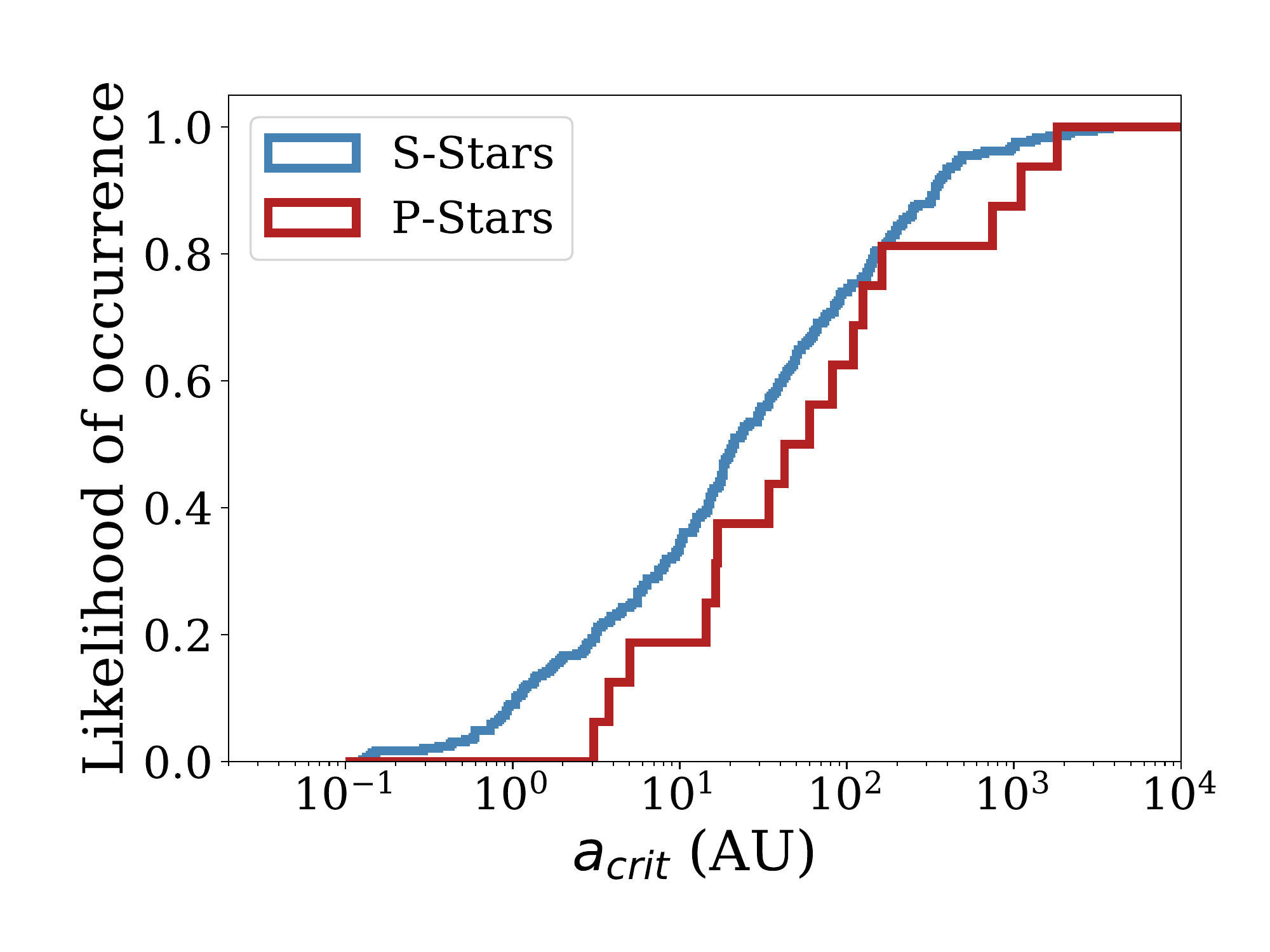}
    \caption{\textbf{Left:} Fraction of planet-host binaries as a function of $a_{crit}$. \textbf{Right:} Cumulative distribution of the $a_{crit}$ for UD pairs with (red) and without (blue) planetary companions. }
    \label{fig:barplot}
\end{figure}

Thanks to the improvement in the information on both companion masses and orbital characteristics, especially for the closest pairs, we were able to better characterize the planet frequencies for these systems by splitting the $a_{crit}<20 au$ bin from BD07 into three smaller bins. This allowed us to check the planet occurrence in binaries with $a_{crit}$ comparable to the baseline used for the definition of the completeness of the UD sample, 2.5~au (see \citep[][]{FV05} for details).
The results shown in Figure~\ref{fig:barplot} seem to suggest that the planet frequency, which is low for very close pairs, rapidly increases and remains stable for higher separations.

The lack of planets in the innermost bin is strong evidence of the negative effects of very close
companions for planets around the components. 
A binary system with $a_{crit}<$ 2.5 au is expected to have undergone a truncation of the circumstellar disk to separations below the snowline, most likely preventing the formation of giant planets (see, e.g., \citep[][]{pichardo2005}).
The observed null result in this range is therefore not surprising.
The run of planet frequency with $a_{crit}$ at larger separations needs additional considerations, as it somewhat depends on the adopted binning $a_{crit}$.
With the binning shown in the left panel of Figure \ref{fig:barplot}, the planet frequency results are similar overall.
However, the cumulative distribution, shown in the right panel of the same figure, highlights an irregular run of $a_{crit}$ for stars with planets, with several plateaus seemingly suggesting a lack of planets in some separation ranges.
Three out of 25 binaries with $a_{crit}$ between to 3 to 5.9 au host planets, though none of the 31 pairs with $a_{crit}$ between 6 to 14 au host planets. This may be due to the small-number statistics and partially due to some remaining ambiguity in the $a_{crit}$ value for binaries without direct detection of the companions.\vspace{-12pt}

One should also note that a value of $a_{crit}$ between 6 to 14 au corresponds to physical separations
between the components of the order of 100~au.
According to \cite{duchene2010}, this represents the distance at which the effect of the presence of the secondary on the protoplanetary disk starts to become important. They suggest that a companion closer than 100~au would alter the properties of the disk, causing disk fragmentation to be the dominant process of planet formation and resulting in a higher fraction of high-mass planets in these systems. If indeed the planet formation process acting in this kind of binary is
more effective at closer separations, this could explain the observed bi-modal distribution. 
Unfortunately, the small number of objects considered and the ambiguities in the determination of $a_{crit}$ render these conclusions merely tentative. Additional observational efforts aimed both at mass and orbit determinations for these systems
will allow more robust assessment.  

\textls[-15]{The fact that the planet frequency of planets around components of wide binaries and single stars is very similar agrees with this overall picture, where one would expect that wide companions do not have any significant effect on the disk properties. 
The only exception to this behavior in the cumulative plot is the lack of planets in binaries with  
$a_{crit}$ between to 200 to 700 au, corresponding to physical separations of a few thousands of au. 
Any impact of the presence of a companion at such large separations is hard to explain from a theoretical perspective, 
and may be due to small-number statistics. }

\begin{table}[H]
\caption{Frequency of planets in binaries with different values of  $a_{crit}$.}\label{tab:freq}
\centering
\begin{tabular}{cccc}
\toprule
  \boldmath{$a_{crit}$} & \boldmath{$N_{star}$} & \boldmath{$N_{planets}$} & \boldmath{$N_{planets}/N_{stars}$}\\
\midrule
 $<$2.5 au   & 57  &  0  &  \\
 2.5--10 au & 50  &  3  &  0.060 $\pm$ 0.0431 \\
  10--20 au & 47  &  3  &  0.064 $\pm$ 0.0462 \\
   20--50 au & 44  &  2  &  0.045 $\pm$ 0.0390 \\
 50--100 au & 33  &  2  &  0.061 $\pm$ 0.0534 \\
100--250 au & 39  &  3  &  0.077 $\pm$ 0.0567 \\
  $>$ 250 au & 43  &  3  &  0.070 $\pm$ 0.0509  \\
\midrule
         UD Singles sub-sample & 537 & 34 & 0.063 $\pm$ 0.0136 \\
   Entire UD binary sub-sample & 313 & 16 & 0.051 $\pm$ 0.0157 \\
\bottomrule
\end{tabular}
\end{table}

\subsection{Stars with White Dwarfs Companions}
\label{sec:wd}

Our extensive census of binary companions allows us to consider the specific case of stars with white dwarf companions known as Sirius-like systems (see \citep[][]{holberg2013} for details).
Ten systems belonging to this category are identified in the sample, including three which were
previously unknown\footnote{Three additional UD objects have white dwarf companions at separations wider than another closer companion, whose dynamical effect is then dominant over that of the white dwarf.}. Planets fulfilling the UD definition are found around two of these stars (HD 13445 and HD 27442). 
This would indicate a rather large frequency of planets in these systems (20\%), which is surprising given that the presence of a white dwarf companion implies an originally tighter binary configuration, due to the mass loss from the 
white dwarf progenitor \citep{DesideraBarbieri2007}.
On the other hand, a larger planet frequency for stars with white dwarf companions may be ascribed to accretion of material lost by the white dwarfs on low-mass planets existing around the companion, making them detectable with the RV technique, or even the formation of second-generation planets \citep{perets2010}.
These scenarios are quite speculative, and~any claims that the observed higher frequency is real are definitely premature.
Our census of white dwarf companions is, in fact, most likely incomplete, as they are typically faint and may also escape correct classifications when detected in one photometric band only (about a dozen of objects from Gaia) or only by dynamical signatures.
Moreover, we expect the original selection biases of the RV planet surveys to have also caused the
exclusion of white dwarf companions with short cooling ages when close enough to have caused spin-up of the companions
(see, e.g., \citep[][]{zurlo2013}).

\section{Summary and Conclusions}
\label{sec:conclusions}

 This paper presents a new, and long overdue, census of the binarity of the stars in the Uniform Detectability (UD) sample by FV05, a widely used reference sample for statistics of giant planets detected via the radial velocity method. 
Building upon the work presented in BD07, we extended the search for binary companions to the stars in the Uniform Detectability (UD) sample by FV05, and were able to add 114 new pairs to the 199 included in the original UD binary sub-sample. 
The information made available in the past few years, in particular thanks to the second Gaia Data Release \cite{gaia_dr2}, allowed us to partially overcome the incompleteness affecting previous studies, which was mainly due to the lack of information on the binary population in most RV survey samples, both in terms of detection and orbital characterization.

In the updated sample, the level of completeness for binaries with separations $>2^{\prime\prime}$, and therefore not excluded due to the selection biases affecting the UD input lists, is $\sim 95\%$.  
As a result, the new value of the overall frequency of binaries in the UD sample is $\sim 37\%$, which is closer to, although still lower that, what is expected for a volume-limited sample---57\%, according to \citep[][]{DM91}.

We used this larger and improved UD binary sample to perform a more unbiased statistical analysis, much less affected by the incompleteness that characterized the original work. Our final goal was to assess the nature and extent of the influence of stellar multiplicity on the formation and, consequently, on the frequency of planets.  
Given the high level of precision achieved in the estimation of the system's parameters, we were able to attempt a characterization of the behavior of the planet frequency among several different sub-sets of binaries in the updated sample.
For this analysis, we~chose to use the critical semi-major axis ($a_{crit}$) defined by \cite{HolmanWiegert1999} as a reference, as it allows one to better take into account the dynamical effects of the presence of the companion on the circumstellar region as well as on planet formation and stability. 

Similarly to BD07, we did not find any evidence of statistically significant differences in the overall planet occurrence between the UD binaries and single stars (5.1 $\pm$ 1.57\% and 6.3 $\pm$ 1.36\%, respectively). 
The lack of planets in close systems ($a_{crit}$ < 2.5 au) found by BD07 was also confirmed, once again reinforcing the expectation of the effect of the presence of a companion on the tidal truncation of the disk.
We also observed an increase of the frequency for systems with $3 < a_{crit} < 6$ au, followed by a decrease for $6 < a_{crit} < 14$ au. This apparent bi-modal distribution, if confirmed, could point towards a positive effect of the binarity on the planet formation process. This could also imply that disk fragmentation is more effective than core accretion in these environments, thanks to the altered properties of the disk caused by these kinds of stellar companions. 
Finally, we found a relatively high value of the planet frequency in the subset of stars hosting white dwarf companions. Several scenarios could be responsible for this, although we feel that it is premature to speculate on the causes without being certain of the authenticity of the result. 

Unfortunately, given the small number of systems considered and the uncertainties still affecting the estimation of the $a_{crit}$, most of the observed trends still need clarification and will certainly benefit from future updates to the information on the binary population, which will soon be available through future Gaia data releases as well as new dedicated observing campaigns. 
Nevertheless, our results represent a further confirmation that planets can form in binary systems in spite of the unfavorable conditions, and the observed trends, if confirmed, could point towards the need for a different formation scenario to explain the observed behavior of the frequency, especially in the case of very tight binaries.

\vspace{6pt} 



\authorcontributions{Data curation, M.B. and S.D.; Investigation, M.B. and S.D.; Methodology, M.B.; Writing---original draft, M.B.; Writing---review and editing, S.D. All authors have read and agreed to the published version of the manuscript.}

\funding{M.B. acknowledges funding by the UK Science and Technology Facilities Council (STFC) grant no. ST/M001229/1. S.D. acknowledges the support by INAF/Frontiera through the “Progetti Premiali” funding scheme of the Italian Ministry of Education, University, and Research.}

\acknowledgments{We thank the anonymous referees for their extensive feedback that significantly improved the clarity of the paper. 
This work made use of data from the European Space Agency (ESA) mission {\it Gaia} (\url{https://www.cosmos.esa.int/gaia}), processed by the {\it Gaia} Data Processing and Analysis Consortium (DPAC, \url{https://www.cosmos.esa.int/web/gaia/dpac/consortium}). Funding for the DPAC has been provided by national institutions, particularly the institutions participating in the {\it Gaia} Multilateral Agreement. 
This research made use of the SIMBAD database, operated at CDS, Strasbourg, France. }

\conflictsofinterest{The authors declare no conflict of interest.
} 

%
%


\appendixtitles{yes} 
\appendix
\section{The UD Binary Sample}
\vspace{-12pt}

\begin{table}[H]
\centering 
\caption{\label{tab:binary_UD}  Properties of binaries found in the UD sample.
The new companions are marked with \textit{N}. 
If the companion was retrieved in Gaia DR2 \citep{gaia_dr2}, the coordinates of the pair were used to update the projected separation ($rho$). 
If an orbital solution was available, the appropriate semi-major axis ($a~(au)$) was included, together with the value of the eccentricity ($ecc$) and of the masses of both the primary ($M_A$) and the companions ($M_B$), if available.
For systems for which only the projected separation was available (empty spaces in the eccentricity column), the semi-major axis was derived from the $rho$ using the relation $a~(au)=1.0~rho~(arcsec)~d(pc)$.
The last two columns include a flag reporting notes for the object from BD07 ($F_{BD07}$:) and any additional notes arising from the present update. A dedicated entry can be found in the appendix for the objects marked with \textbf{*} in the last column.  
The mass flag indicates the source for the companion mass:  \textbf{a:} $M_{comp}$ from VF06;  \textbf{b:} $M_{comp}$ from \mbox{\citet{1997AJ....113.2246R,Delfosse2000}}; \textbf{ c:} $M_{comp}$ from individual papers (see Reference list below), \textbf{ d:} minimum $M_{comp}$ compatible with $\Delta\mu$ and/or RV trend and null detection in GDR2, and \textbf{e:} preliminary orbital solution based on available RV measurements.}
\scalebox{0.85}[0.85]{
\begin{tabular}{lccccccclll}
\toprule
\multicolumn{11}{l}{Stars with planets (as defined in FV05)}\\ 
\midrule
\textbf{ID}          &   \textbf{plx  (mas)}&    \textbf{ rho (\boldmath{$\prime\prime$}) }	&  \textbf{ecc}      &   \textbf{a (au)}     & \textbf{a\boldmath{$_{crit}$} (au)} & \textbf{M\boldmath{$_A$}  M\boldmath{$_{\odot}$} }     & \textbf{M\boldmath{$_B$}   M\boldmath{$_{\odot}$}}     & \textbf{M\boldmath{$_{Flag}$}} & \textbf{BD07 }     & \textbf{Notes} \\
\midrule
HD142 & 38.16 & 3.89 &  &  & 14.79 & 1.24 & 0.53 & b &  & $\Delta\mu$\\
HD9826 & 74.57 & 55.62 &  &  & 125.92 & 1.32 & 0.19 & b &  & \\
HD13445 & 92.7 &  & 0.4 & 18.4 & 3.07 & 0.77 & 0.49 & c & $\Delta\mu$ (*) & $\Delta\mu$ RV \\
HD20782 & 27.76 & 252.99 &  &  & 1127.2 & 1.0 & 0.84 & c & (*) &  \\
HD27442 & 54.71 & 13.04 &  &  & 34.99 & 1.49 & 0.6 & c & (*) &  \\
HD38529 & 23.58 & 283.65 &  &  & 1839.64 & 1.47 & 0.47 & b & G(*) &  \\
HD40979 & 29.31 & 191.38 &  &  & 765.14 & 1.19 & 1.21 & c & (*) & $\Delta\mu$(*) \\ 
HD46375 & 33.81 & 10.44 &  &  & 42.4 & 0.92 & 0.51 & b &  & $\Delta\mu$\\
HD75732 & 79.43 & 84.82 &  &  & 166.23 & 0.91 & 0.26 & b &  & \\
HD120136 & 63.86 &  & 0.87 & 221.0 & 5.13 & 1.35 & 0.49 & c & (*) &  \\
HD 177830 & 15.9 & 1.6 &  &  & 16.86 & 1.47 & 0.23 & c &  & N \\
HD178911B & 24.38 & 13.6 &  &  & 60.28 & 1.42 & 1.89 & c & (*) & $\Delta\mu$ \\
HD188015 & 19.71 & 13.03 &  &  & 111.06 & 1.25 & 0.19 & b &  & \\ \bottomrule
\end{tabular}}  
\end{table}

\begin{table}[H]\ContinuedFloat
\centering \small
\caption{{\em Cont.}} 
\label{tab:binary_UD}
\scalebox{0.85}[0.85]{
\begin{tabular}{lccccccclll}
\toprule
\multicolumn{11}{l}{Stars with planets (as defined in FV05)}\\\midrule
\textbf{ID}          &   \textbf{plx  (mas)}&    \textbf{ rho (\boldmath{$\prime\prime$}) }	&  \textbf{ecc}      &   \textbf{a (au)}     & \textbf{a\boldmath{$_{crit}$} (au)} & \textbf{M\boldmath{$_A$}  M\boldmath{$_{\odot}$} }     & \textbf{M\boldmath{$_B$}   M\boldmath{$_{\odot}$}}     & \textbf{M\boldmath{$_{Flag}$}} & \textbf{BD07 }     & \textbf{Notes} \\
\midrule
HD195019 & 26.6 & 3.39 &  &  & 17.32 & 1.07 & 0.62 & b &  & \\
HD196050 & 19.71 & 10.77 &  &  & 83.85 & 1.15 & 0.36 & b &  & (*) \\
HD196885 & 29.24 &  & 0.42 & 21.0 & 3.8 & 1.25 & 0.45 & c & (*) & $\Delta\mu$(*) \\
HD222582 & 23.69 & 113.3 &  &  & 715.48 & 0.99 & 0.36 & b &  & \\
\midrule
\multicolumn{11}{l}{ Stars without planets (as defined in FV05)}\\ 
\midrule
\textbf{ID}          &   \textbf{plx  (mas)}&    \textbf{ rho (\boldmath{$\prime\prime$}) }	&  \textbf{ecc}      &   \textbf{a (au)}     & \textbf{a\boldmath{$_{crit}$} (au)} & \textbf{M\boldmath{$_A$}  M\boldmath{$_{\odot}$} }     & \textbf{M\boldmath{$_B$}   M\boldmath{$_{\odot}$}}     & \textbf{M\boldmath{$_{Flag}$}} & \textbf{BD07 }     & \textbf{Notes} \\
\midrule
HD531A & 14.21 & 5.23 &  &  & 43.49 & 1.66 & 1.64 & a &  & $\Delta\mu$\\
HD531B & 14.1 & 5.23 &  &  & 43.52 & 1.64 & 1.66 & a &  & $\Delta\mu$\\
HD1388 & 37.11 &  &  & 106.85 & 18.43 & 1.18 & 0.13 & d &  & N  $\Delta\mu$\\
HD3074 & 29.46 & 4.8 &  &  & 21.15 & 1.2 & 0.84 & b &  & \\
HD3079 & 20.81 & 3.43 &  &  & 24.23 & 1.05 & 0.42 & b &  & N  $\Delta\mu$ \\
HD3651 & 89.79 & 43.2 &  &  & 85.34 & 0.89 & 0.06 & b & (*) &  \\
HD3770 & 14.97 & 0.3 &  &  & 2.79 & 1.25 & 0.65 & b & S RV & $\Delta\mu$ \\
HD3795 & 35.13 &  &  & 92.02 & 10.79 & 1.94 & 1.97 & d & S RV $\Delta\mu$ & $\Delta\mu$ \\
HD3821 & 37.74 & 8.45 &  &  & 29.72 & 1.0 & 0.64 & b &  & \\
HD3823 & 39.6 & 7.31 &  &  & 30.56 & 1.25 & 0.22 & b &  & N \\
HD4307 & 30.81 & 4.08 &  &  & 21.64 & 1.31 & 0.26 & b &  & N \\
HD4614 & 168.75 &  & 0.49 & 72.0 & 10.0 & 0.99 & 0.58 & b &  & \\
HD4747 & 53.18 &  & 0.74 & 10.01 & 0.75 & 0.82 & 0.07 & c & SB & $\Delta\mu$\\
HD4903 & 18.05 &  &  & 84.73 & 15.47 & 1.17 & 0.03 & d &  & N  $\Delta\mu$\\
HD5470 & 14.85 &  &  & 20.5 & 3.24 & 1.13 & 0.29 & c &  & $\Delta\mu$\\
HD6558 & 12.18 & 19.02 &  &  & 253.26 & 1.23 & 0.26 & b &  & N  $\Delta\mu$ \\
HD6734 & 21.37 &  &  & 80.0 & 13.99 & 2.25 & 0.2 & a & $\Delta\mu$ & $\Delta\mu$  RV \\
HD6872A & 11.08 & 14.51 &  &  & 180.95 & 1.91 & 1.03 & a &  & \\
HD6872B & 11.04 & 14.51 &  &  & 127.91 & 1.03 & 1.91 & a &  & \\
HD7963 & 8.55 &  & 0.04 & 23.4 & 5.98 & 0.84 & 0.89 & b & (*) &  \\
HD8648 & 25.04 &  &  & 105.76 & 19.19 & 1.16 & 0.04 & d &  & N  $\Delta\mu$\\
HD8673 & 26.38 &  & 0.5 & 48.5 & 7.55 & 1.36 & 0.39 & c &  & N  $\Delta\mu$(*) \\
HD8765 & 20.47 &  & 0.0 & 5.3 & 0.95 & 1.2 & 0.06 & c & G  $\Delta\mu$ & $\Delta\mu$(*) \\
HD9331 & 18.01 & 16.33 &  &  & 152.46 & 0.93 & 0.14 & b &  & N \\
HD10360 & 122.06 &  & 0.53 & 52.2 & 5.66 & 0.75 & 0.77 & a &  & $\Delta\mu$\\
HD10361 & 122.13 &  & 0.53 & 52.2 & 5.74 & 0.77 & 0.75 & a &  & $\Delta\mu$ \\
HD11112 & 22.5 & 2.2 &  &  & 12.56 & 1.37 & 1.0 & c &  & N  $\Delta\mu$  (*) \\
HD11964 & 29.79 & 29.68 &  &  & 138.66 & 1.13 & 0.59 & b & (*) &  \\
HD12414 & 21.49 & 9.64 &  &  & 60.59 & 0.84 & 0.50 & c &  & N (*) \\
HD13043 & 26.92 & 79.21 &  &  & 400.02 & 1.14 & 0.66 & b &  & \\
HD13612 B & 26.09 & 16.76 &  &  & 58.58 & 1.02 & 2.32 & c & (*) & $\Delta\mu$\\

HD13507 & 38.18 &  & 0.38 & 8.54 & 1.79 & 1.14 & 0.28 & c & SB  $\Delta\mu$ (*) & $\Delta\mu$ \\
HD13531 & 38.27 & 0.7 &  &  & 2.98 & 0.94 & 0.19 & c &  & $\Delta\mu$ \\
HD13825 & 38.89 &  &  & 105.13 & 17.47 & 1.18 & 0.2 & d &  & N  $\Delta\mu$ \\
HD16141 & 26.44 & 6.27 &  &  & 37.06 & 1.15 & 0.32 & b & (*) &  \\
HD16160 & 138.21 &  & 0.75 & 15.0 & 1.04 & 0.76 & 0.09 & c & (*) &  \\
HD16417 & 39.35 & 45.01 &  &  & 201.23 & 1.38 & 0.11 & b &  & N \\
HD16287 & 41.09 &  &  & 169.33 & 31.03 & 0.97 & 0.02 & d &  & N  $\Delta\mu$\\
HD16895 & 89.7 &  & 0.13 & 249.5 & 74.23 & 1.24 & 0.49 & b &  & \\
HD18143 & 44.37 & 6.5 &  &  & 18.57 & 0.9 & 0.69 & b & (*) &  \\
HD18445 & 40.26 &  & 0.56 & 1.06 & 0.14 & 0.78 & 0.18 & c & S(*) & $\Delta\mu$ \\
HD18907 & 31.25 &  & 0.28 & 13.0 & 3.86 & 2.04 & 0.02 & c &  & N  SB  $\Delta\mu$(*) \\
HD19467 & 31.23 & 1.65 &  &  & 9.54 & 1.45 & 0.06 & c &  & N  SB  $\Delta\mu$(*) \\
HD20766 & 83.06 & 310.0 &  &  & 405.51 & 0.91 & 1.19 & a &  & \\
HD20807 & 83.01 & 310.0 &  &  & 418.8 & 0.95 & 1.12 & a &  & \\
HD21019 & 26.23 & 3.9 &  &  & 20.31 & 1.11 & 0.63 & b &  & \\
HD22879 & 38.53 & 19.19 &  &  & 87.07 & 1.12 & 0.1 & b &  & N  $\Delta\mu$ \\
HD23439 & 41.83 & 6.89 &  &  & 18.0 & 0.67 & 0.86 & c & (*) & $\Delta\mu$ \\
HD24496 & 48.81 & 3.55 &  &  & 10.43 & 1.08 & 0.49 & b &  & N  $\Delta\mu$ \\
HD25682 & 23.9 & 10.16 &  &  & 65.47 & 1.34 & 0.41 & b &  & N  $\Delta\mu$ \\
HD26491 & 42.46 &  & 0.57 & 10.0 & 1.18 & 1.06 & 0.5 & c & $\Delta\mu$ & SB (*) \\
HD28255A & 36.18 & 5.34 &  &  & 17.43 & 1.07 & 1.06 & a &  & \\
HD28255B & 36.26 & 5.34 &  &  & 17.29 & 1.06 & 1.07 & a &  & $\Delta\mu$ \\
HD29461 & 22.12 &  & 0.0 & 7.3 & 1.25 & 1.2 & 0.15 & c & RV & $\Delta\mu$(*) \\  
HD29836 & 23.51 & 141.71 &  &  & 682.33 & 1.19 & 1.36 & c & (*) &  \\
HD30339 & 14.74 &  & 0.25 & 0.13 & 0.04 & 1.39 & 0.07 & c & SB & $\Delta\mu$\\ 
HD30649 & 30.88 &  & 0.6 & 16.5 & 1.99 & 0.9 & 0.21 & e & RV (*)& $\Delta\mu$ \\
HD31253 & 17.17 & 416.02 &  &  & 3020.57 & 1.25 & 1.02 & b &  & N \\
HD31412 & 27.74 & 0.2 &  &  & 1.08 & 1.17 & 0.43 & c & RV (*) & $\Delta\mu$ (*) \\
HD31966 & 26.53 & 10.51 &  &  & 67.59 & 1.26 & 0.16 & b &  & N \\
HD32923 & 62.95 &  & 0.9 & 2.86 & 0.03 & 1.03 & 1.11 & b &  & $\Delta\mu$\\ \bottomrule
\end{tabular}}  \end{table}

\begin{table}[H]\ContinuedFloat
\centering \small
\caption{{\em Cont.}} 
\label{tab:binary_UD}
\scalebox{0.85}[0.85]{
\begin{tabular}{lccccccclll}
\toprule
\textbf{ID}          &   \textbf{plx  (mas)}&    \textbf{ rho (\boldmath{$\prime\prime$}) }	&  \textbf{ecc}      &   \textbf{a (au)}     & \textbf{a\boldmath{$_{crit}$} (au)} & \textbf{M\boldmath{$_A$}  M\boldmath{$_{\odot}$} }     & \textbf{M\boldmath{$_B$}   M\boldmath{$_{\odot}$}}     & \textbf{M\boldmath{$_{Flag}$}} & \textbf{BD07 }     & \textbf{Notes} \\
\midrule
HD33473 & 18.19 & 10.28 &  &  & 76.93 & 1.32 & 0.76 & b &  & $\Delta\mu$ \\
HD33632 & 37.65 & 33.99 &  &  & 144.93 & 1.09 & 0.25 & b &  & N \\
HD33636 & 33.72 &  & 0.48 & 3.4 & 0.61 & 1.02 & 0.14 & c &  & N  $\Delta\mu$(*) \\
HD35681 & 29.17 & 10.13 &  &  & 54.9 & 1.05 & 0.27 & b &  & N \\
HD35956 & 33.72 &  & 0.62 & 2.6 & 0.3 & 0.98 & 0.18 & c & SB(*) &  \\
HD37394 & 81.43 & 98.03 &  &  & 163.55 & 0.93 & 0.54 & b &  & \\
HD39213 & 16.37 &  & 0.2 & 2.3 & 0.75 & 0.97 & 0.07 & c &  & N  SB  $\Delta\mu$ (*)\\
HD39587 & 113.12 &  & 0.45 & 5.9 & 1.13 & 1.05 & 0.14 & c & SB & $\Delta\mu$\\
HD39881 & 36.34 & 47.78 &  &  & 233.68 & 1.55 & 0.1 & b &  & N \\
HD40397 & 41.09 & 2.23 &  &  & 7.49 & 0.92 & 0.5 & b &  & $\Delta\mu$ \\
HD40650 & 36.05 &  &  & 96.38 & 15.17 & 1.02 & 0.27 & d &  & N  $\Delta\mu$\\
HD42024 & 16.8 &  & 0.19 & 0.38 & 0.13 & 1.35 & 0.07 & c &  & N  SB (*) \\
HD43834 & 97.9 & 3.0 &  &  & 5.18 & 0.98 & 0.14 & c &  & N \\
HD44120 & 27.66 & 40.8 &  &  & 212.93 & 1.23 & 0.54 & c &  & $\Delta\mu$  (*) \\
HD43587 & 51.8 &  & 0.8 & 11.6 & 0.53 & 1.06 & 0.34 & c & SB (*) & $\Delta\mu$ \\
HD44985 & 29.91 & 9.29 &  &  & 49.14 & 0.9 & 0.23 & b &  & N \\
HD45588 & 34.14 & 41.3 &  &  & 175.38 & 1.21 & 0.52 & b &  & \\
HD45701 & 31.3 &  & 0.17 & 22.28 & 5.12 & 1.18 & 1.0 & c & $\Delta\mu$ & SB  $\Delta\mu$(*) \\
HD47157 & 24.77 & 10.7 &  &  & 61.76 & 1.13 & 0.52 & b &  & \\
HD50281 & 114.3 & 58.3 &  &  & 63.32 & 0.76 & 0.63 & b &  & (*) \\
HD50639 & 25.68 &  & 0.0 & 20.0 & 3.41 & 1.16 & 0.15 & d & RV  $\Delta\mu$ & \\
HD51929 & 26.71 & 0.72 &  &  & 3.93 & 0.86 & 0.36 & c & $\Delta\mu$ & $\Delta\mu$(*) \\
HD52447 & 12.51 &  &  & 36.27 & 6.2 & 1.14 & 0.14 & d &  & N  RV, $\Delta\mu$(*) \\
HD53705 & 58.64 & 21.25 &  &  & 43.73 & 0.97 & 0.89 & a & (*) &  \\
HD53706 & 58.64 & 21.25 &  &  & 41.61 & 0.89 & 0.97 & a &  & (*) \\
HD61606 & 70.92 & 57.9 &  &  & 103.54 & 0.81 & 0.62 & b &  & \\
HD63754 & 19.88 & 5.61 &  &  & 42.29 & 1.5 & 0.54 & b &  & $\Delta\mu$\\
HD64184 & 29.94 &  & 0.24 & 0.13 & 0.04 & 1.41 & 0.17 & c &  & N  SB \\
HD64468 & 50.34 &  & 0.26 & 0.56 & 0.15 & 0.81 & 0.14 & c & SB & \\
HD65277 & 56.65 & 5.13 &  &  & 12.72 & 0.72 & 0.36 & c & $\Delta\mu$ & $\Delta\mu$ (*) \\
HD65430 & 42.54 &  & 0.32 & 4.0 & 1.05 & 0.83 & 0.06 & c & SB & $\Delta\mu$\\
HD65907 & 61.73 & 58.62 &  &  & 135.93 & 1.77 & 0.81 & c & (*) &  \\
HD67458 & 38.86 & 8.82 &  &  & 38.34 & 1.05 & 0.15 & b &  & N  $\Delta\mu$\\
HD66171 & 21.14 & 49.3 &  &  & 347.18 & 0.91 & 0.34 & b &  & $\Delta\mu$\\
HD68017 & 46.33 &  &  & 61.0 & 8.47 & 1.55 & 0.82 & d &  & N  $\Delta\mu$\\
HD71881 & 23.86 &  &  & 115.92 & 21.41 & 1.01 & 0.01 & d &  & N  $\Delta\mu$\\
HD72760 & 47.4 & 0.9 &  &  & 3.21 & 0.91 & 0.13 & c & $\Delta\mu$ (*) & $\Delta\mu$\\

HD72780 & 19.27 &  & 0.0 & 9.3 & 1.68 & 1.28 & 0.05 & c & RV & $\Delta\mu$(*) \\
HD73344 & 28.33 &  &  & 132.01 & 24.41 & 1.26 & 0.01 & d &  & N  $\Delta\mu$\\
HD73668 & 28.07 &  & 0.0 & 8.0 & 1.35 & 1.13 & 0.17 & c &  & $\Delta\mu$ (*) \\
HD74014 & 28.74 &  & 0.53 & 7.22 & 1.2 & 1.04 & 0.05 & c &  & N  SB  $\Delta\mu$ \\
HD76752 & 26.14 &  &  & 68.28 & 10.63 & 1.05 & 0.3 & d &  & N  $\Delta\mu$ \\
HD77407 & 29.53 & 1.72 &  &  & 8.25 & 1.12 & 0.54 & c & (*) & $\Delta\mu$  \\
HD80913 & 16.27 &  & 0.0 & 71.6 & 12.88 & 1.27 & 0.06 & d &  & N  RV $\Delta\mu$(*) \\
HD86264 & 14.72 & 46.6 &  &  & 493.8 & 1.88 & 0.53 & b &  & N \\
HD86728 & 67.0 & 134.38 &  &  & 343.64 & 1.08 & 0.13 & b & (*) & \\
HD87424 & 42.94 & 9.73 &  &  & 38.02 & 0.78 & 0.12 & b &  & N \\
HD87359 & 31.95 &  &  & 77.79 & 12.05 & 1.05 & 0.31 & d &  & N  $\Delta\mu$ \\
HD88218 & 31.5 & 2.0 &  &  & 8.45 & 1.09 & 0.69 & c &  & RV  $\Delta\mu$ \\
HD90839 & 77.43 & 122.86 &  &  & 217.96 & 1.12 & 0.62 & b & (*) &  \\
HD92222 & 14.48 & 18.1 &  &  & 148.78 & 1.09 & 1.05 & b & (*) &  \\
HD93385 & 23.04 & 10.39 &  &  & 67.8 & 1.16 & 0.41 & b &  & N  $\Delta\mu$\\
HD93745 & 16.94 & 435.71 &  &  & 3730.8 & 1.12 & 0.48 & b &  & N \\
HIP52942 & 6.07 & 17.52 &  &  & 322.58 & 1.04 & 1.24 & b & (*) & $\Delta\mu$ \\
HIP52940 & 5.72 &  & 0.37 & 2.6 & 0.6 & 1.12 & 0.12 & c & SB(*) &  \\
HD96574 & 20.04 &  &  & 89.08 & 16.18 & 1.21 & 0.04 & d &  & N  $\Delta\mu$\\
HD97037 & 30.82 &  &  & 116.9 & 20.79 & 1.26 & 0.08 & d &  & N  $\Delta\mu$\\
HD97334 & 44.14 & 89.88 &  &  & 366.55 & 1.09 & 0.05 & c & (*) & $\Delta\mu$ \\
HD98618 & 24.24 & 2.7 &  &  & 15.99 & 1.04 & 0.47 & b &  & N  $\Delta\mu$\\
HD99491 & 54.92 & 28.18 &  &  & 63.23 & 1.01 & 0.86 & a &  & \\
HD99492 & 54.91 & 28.18 &  &  & 57.62 & 0.86 & 1.01 & a &  & (*) \\
HD100180 & 42.26 & 15.5 &  &  & 48.92 & 1.1 & 0.69 & b &  & \\ 
HD100623 & 104.74 & 15.3 &  &  & 17.97 & 0.77 & 0.66 & c &  & RV $\Delta\mu$  (*) \\ 
HD101177 & 42.79 & 9.7 &  &  & 26.24 & 0.99 & 1.05 & c & (*) &  \\ 
HD101259 & 14.76 & 7.99 &  &  & 93.36 & 2.27 & 0.25 & b &  & N  $\Delta\mu$\\
HD102365 & 107.62 & 22.73 &  &  & 34.29 & 0.86 & 0.18 & b &  & $\Delta\mu$\\
HD103432 & 25.24 & 73.41 &  &  & 345.76 & 0.92 & 0.89 & b &  & \\
HD103829 & 9.93 &  & 0.27 & 6.2 & 1.75 & 1.2 & 0.13 & b & RV & $\Delta\mu$\\
HD105113 & 20.65 & 5.74 &  &  & 31.16 & 1.28 & 1.51 & c &  & (*) \\ 
HD106453 & 34.03 & 4.3 &  &  & 18.04 & 0.82 & 0.38 & b &  & N  $\Delta\mu$ \\\bottomrule
\end{tabular}}  \end{table}

\begin{table}[H]\ContinuedFloat
\centering \small
\caption{{\em Cont.}} 
\label{tab:binary_UD}
\scalebox{0.85}[0.85]{
\begin{tabular}{lccccccclll}
\toprule
\textbf{ID}          &   \textbf{plx  (mas)}&    \textbf{ rho (\boldmath{$\prime\prime$}) }	&  \textbf{ecc}      &   \textbf{a (au)}     & \textbf{a\boldmath{$_{crit}$} (au)} & \textbf{M\boldmath{$_A$}  M\boldmath{$_{\odot}$} }     & \textbf{M\boldmath{$_B$}   M\boldmath{$_{\odot}$}}     & \textbf{M\boldmath{$_{Flag}$}} & \textbf{BD07 }     & \textbf{Notes} \\
\midrule
HD107148 & 20.21 & 34.98 &  &  & 253.38 & 1.37 & 0.66 & c &  & N \\
HD107705 & 33.32 & 20.96 &  &  & 85.42 & 1.22 & 0.71 & b &  & $\Delta\mu$\\
HD107692 & 38.41 & 17.79 &  &  & 81.71 & 1.06 & 0.08 & b &  & N  $\Delta\mu$\\
HD109358 & 116.13 &  &  & 293.77 & 45.5 & 1.05 & 0.31 & d &  & N  $\Delta\mu$\\
HD110810 & 50.49 & 21.81 &  &  & 72.98 & 0.77 & 0.11 & b &  & N  $\Delta\mu$\\
HD111398 & 27.5 & 75.73 &  &  & 452.54 & 1.06 & 0.2 & b &  & \\
HD111484A & 11.45 & 9.03 &  &  & 92.67 & 1.38 & 1.39 & a &  & \\
HD111484B & 11.45 & 9.03 &  &  & 93.06 & 1.39 & 1.38 & a &  & \\
HD114174 & 37.91 &  & 0.77 & 33.4 & 1.51 & 1.34 & 1.2 & c &  & N  $\Delta\mu$  (*) \\
HD114729 & 26.42 & 8.19 &  &  & 46.92 & 1.0 & 0.34 & b & (*) &  \\
HD114853 & 39.92 & 35.34 &  &  & 149.43 & 1.04 & 0.15 & b &  & N \\
HD116442 & 60.25 & 26.2 &  &  & 51.8 & 0.76 & 0.73 & a &  & \\
HD116443 & 60.34 & 26.2 &  &  & 50.53 & 0.73 & 0.76 & a &  & \\
HD117939 & 33.69 & 3.29 &  &  & 14.88 & 1.07 & 0.35 & b &  & N RV (*) \\
HD120066 & 31.76 & 488.5 &  &  & 2092.19 & 1.16 & 0.67 & b &  & \\
HD120237 & 33.22 & 11.65 &  &  & 47.87 & 1.16 & 0.66 & b &  & \\
HD120476 & 73.92 &  & 0.44 & 33.15 & 4.63 & 0.76 & 0.68 & b &  & \\
HD120690 & 53.88 &  & 0.34 & 4.73 & 0.91 & 1.12 & 0.64 & c & $\Delta\mu$ & SB (*) \\
HD120780 & 59.55 & 5.7 &  &  & 13.53 & 0.74 & 0.36 & b & G  $\Delta\mu$ & $\Delta\mu$ (*) \\
HD121384 & 25.69 &  & 0.84 & 0.61 & 0.02 & 1.18 & 0.17 & c &  & SB (*) \\
HD122742 & 60.84 &  & 0.48 & 5.3 & 0.75 & 0.92 & 0.54 & c & SB & \\
HD124694 & 25.44 & 82.73 &  &  & 457.92 & 0.97 & 0.48 & b &  & N \\
HD125455 & 48.6 & 15.18 &  &  & 50.55 & 0.79 & 0.17 & b &  & $\Delta\mu$\\
HD126614 & 13.65 & 0.49 &  &  & 5.64 & 1.19 & 0.32 & c &  & $\Delta\mu$ RV (*) \\
HD128428 & 16.75 & 0.8 &  &  & 6.45 & 1.26 & 0.75 & b &  & RV  $\Delta\mu$\\
HD128621 & 743.0 &  & 0.51 & 22.76 & 2.44 & 0.89 & 1.12 & a & (*) &  \\
HD128620 & 743.0 &  & 0.51 & 22.76 & 2.79 & 1.12 & 0.89 & a & (*) &  \\
HD128674 & 38.92 & 484.73 &  &  & 1661.71 & 0.83 & 0.52 & b &  & \\
HD129814 & 23.81 &  & 0.0 & 30.0 & 4.27 & 1.06 & 0.5 & d & RV $\Delta\mu$ & $\Delta\mu$\\
HD130948 & 54.91 & 2.6 &  &  & 8.22 & 1.11 & 0.11 & c &  & N \\
HD131156 & 148.52 &  & 0.51 & 32.8 & 3.93 & 0.92 & 0.79 & b & (*) &  \\
HD131509 & 12.91 &  &  & 57.99 & 10.5 & 1.33 & 0.05 & d &  & N  $\Delta\mu$\\
HD131511 & 87.91 &  & 0.51 & 0.52 & 0.07 & 0.93 & 0.45 & c & SB & $\Delta\mu$\\
HD131923 & 42.08 &  & 0.28 & 8.7 & 2.02 & 1.04 & 0.42 & c & G  $\Delta\mu$ & SB  $\Delta\mu$ (*) \\
HD131977 & 170.01 & 24.9 &  &  & 15.98 & 0.76 & 0.98 & c & (*) & (*) \\

HD132375 & 29.17 & 9.75 &  &  & 50.11 & 1.28 & 0.46 & b &  & N \\
HD133161 & 26.03 & 0.5 &  &  & 2.86 & 1.18 & 0.44 & b & $\Delta\mu$ & SB  $\Delta\mu$\\
HD134044 & 33.34 & 4.42 &  &  & 20.0 & 0.98 & 0.34 & b &  & N  $\Delta\mu$\\
HD134440 & 33.8 & 300.62 &  &  & 1041.81 & 0.55 & 0.56 & a &  & $\Delta\mu$\\
HD134439 & 33.99 & 300.62 &  &  & 1046.86 & 0.56 & 0.55 & a &  & $\Delta\mu$\\
HD134331 & 31.89 & 49.7 &  &  & 188.48 & 1.12 & 1.02 & a &  & $\Delta\mu$ \\
HD134330 & 31.88 & 49.7 &  &  & 178.6 & 1.02 & 1.12 & a &  & \\
HD135101 & 30.9 & 23.27 &  &  & 92.55 & 1.07 & 0.92 & c &  & \\
HD136118 & 19.41 &  & 0.35 & 2.36 & 0.6 & 1.25 & 0.04 & c & P & N  S $\Delta\mu$\\
HD136442 & 27.52 & 14.03 &  &  & 89.31 & 2.44 & 0.21 & b &  & N \\
HD136580 & 24.6 &  & 0.0 & 20.0 & 3.02 & 1.17 & 0.4 & d & RV  $\Delta\mu$ & $\Delta\mu$ \\
HD137510 & 23.27 &  & 0.4 & 1.88 & 0.44 & 1.39 & 0.03 & c & P & $\Delta\mu$\\
HD137778 & 48.38 & 52.21 &  &  & 106.45 & 0.9 & 1.6 & c & (*) & (*) \\
HD139477 & 51.15 & 42.54 &  &  & 122.26 & 0.75 & 0.3 & b &  & \\
HD139323 & 44.51 & 121.29 &  &  & 270.56 & 0.89 & 1.55 & c & (*) & (*) \\
HD139457 & 20.82 &  &  & 61.31 & 9.79 & 0.88 & 0.21 & d &  & N  $\Delta\mu$ \\
HD140785 & 18.37 & 4.43 &  &  & 38.85 & 1.48 & 0.33 & b &  & N \\
HD140901 & 65.53 & 14.5 &  &  & 30.25 & 1.08 & 0.61 & c &  & (*) \\
HD140913 & 20.67 &  & 0.54 & 0.55 & 0.09 & 1.17 & 0.04 & c & SB & \\
HD141103 & 19.92 & 14.19 &  &  & 122.77 & 0.99 & 0.11 & b &  & N \\
HD140901 & 65.53 & 14.46 &  &  & 29.63 & 1.0 & 0.61 & c &  & $\Delta\mu$  \\
HD142229 & 21.62 &  &  & 6.9 & 1.17 & 1.09 & 0.15 & c & RV & SB  $\Delta\mu$  (*) \\
HD144579 & 69.59 & 70.0 &  &  & 162.95 & 0.75 & 0.16 & b &  & $\Delta\mu$\\
HD144585 & 36.31 & 36.88 &  &  & 180.31 & 1.51 & 0.1 & b &  & N \\
HD145435 & 27.12 & 1.59 &  &  & 9.09 & 1.19 & 0.35 & b & $\Delta\mu$ & $\Delta\mu$ \\
HD145958A & 41.26 &  & 0.39 & 124.0 & 18.66 & 0.9 & 0.89 & a & (*) & $\Delta\mu$\\
HD145958B & 41.26 &  & 0.39 & 124.0 & 18.53 & 0.89 & 0.9 & a & (*) & $\Delta\mu$\\ 
HD145825 & 45.03 &  & 0.34 & 7.3 & 1.82 & 1.08 & 0.1 & c &  & N SB  $\Delta\mu$ (*) \\ 
HD146362 & 44.13 &  & 0.72 & 119.2 & 5.6 & 1.12 & 2.23 & c &  & $\Delta\mu$ (*) \\  
HD146481 & 22.42 & 4.84 &  &  & 34.69 & 1.67 & 0.38 & b &  & N \\
HD147231 & 24.86 &  &  & 73.5 & 12.94 & 1.39 & 0.11 & d &  & N  $\Delta\mu$ \\
HD147723 & 30.56 & 4.08 &  &  & 16.2 & 1.29 & 1.16 & a &  & $\Delta\mu$ \\
HD147722 & 30.56 & 4.08 &  &  & 15.24 & 1.16 & 1.29 & a &  & $\Delta\mu$\\
HD149200 & 19.4 & 179.27 &  &  & 1368.85 & 1.12 & 0.43 & b &  & N \\ 
HD149806 & 49.3 & 6.3 &  &  & 18.84 & 0.94 & 0.37 & b &  & $\Delta\mu$ \\\bottomrule
\end{tabular}}  \end{table}

\begin{table}[H]\ContinuedFloat
\centering \small
\caption{{\em Cont.}} 
\label{tab:binary_UD}
\scalebox{0.85}[0.85]{
\begin{tabular}{lccccccclll}
\toprule
\textbf{ID}          &   \textbf{plx  (mas)}&    \textbf{ rho (\boldmath{$\prime\prime$}) }	&  \textbf{ecc}      &   \textbf{a (au)}     & \textbf{a\boldmath{$_{crit}$} (au)} & \textbf{M\boldmath{$_A$}  M\boldmath{$_{\odot}$} }     & \textbf{M\boldmath{$_B$}   M\boldmath{$_{\odot}$}}     & \textbf{M\boldmath{$_{Flag}$}} & \textbf{BD07 }     & \textbf{Notes} \\
\midrule
HD150248 & 35.99 &  & 0.67 & 4.36 & 0.44 & 0.96 & 0.1 & c & $\Delta\mu$ & SB  $\Delta\mu$(*) \\
HD150554 & 21.16 &  & 0.0 & 7.6 & 1.35 & 1.13 & 0.07 & c &  & $\Delta\mu$ (*) \\
HD150698 & 21.25 & 26.73 &  &  & 220.76 & 1.68 & 0.14 & b &  & N \\
HD151044 & 34.09 &  &  & 161.53 & 29.84 & 1.06 & 0.01 & d &  & N  $\Delta\mu$\\
HD151090 & 22.75 & 163.08 &  &  & 967.04 & 1.17 & 0.7 & b &  & $\Delta\mu$\\
HD151995 & 36.49 &  &  & 107.81 & 19.42 & 0.89 & 0.04 & d &  & N  $\Delta\mu$ \\
HD154160 & 27.45 & 4.08 &  &  & 23.32 & 1.44 & 0.39 & b &  & N \\
HD155060 & 27.31 & 6.5 &  &  & 35.91 & 1.01 & 0.35 & b &  & N  $\Delta\mu$\\
HD156826 & 20.94 & 120.18 &  &  & 984.33 & 2.61 & 0.31 & b &  & N  $\Delta\mu$\\
HD156274 & 120.18 & 0.79 &  &  & 0.85 & 0.79 & 0.55 & b &  & \\
HD156274B & 120.18 &  & 0.0 & 63.5 & 10.24 & 0.9 & 0.2 & c &  & SB (*) \\
HD157338 & 30.18 &  &  & 55.63 & 7.2 & 0.93 & 0.66 & d &  & N  $\Delta\mu$\\
HD158783 & 24.23 &  & 0.05 & 5.0 & 1.92 & 1.15 & 0.19 & c &  & N  SB  $\Delta\mu$(*) \\ 
HD159868 & 17.88 &  &  & 82.58 & 15.04 & 1.39 & 0.04 & d &  & N  $\Delta\mu$\\
HD161797 & 119.11 &  & 0.32 & 22.0 & 5.63 & 1.15 & 0.13 & c & RV  $\Delta\mu$ (*) & RV, $\Delta\mu$ \\
HD162255 & 22.63 &  & 0.06 & 0.27 & 0.1 & 1.34 & 0.33 & c &  & N  SB  $\Delta\mu$(*) \\
HD164595 & 35.36 & 88.12 &  &  & 358.12 & 0.98 & 0.44 & b &  & \\
HD164507 & 22.01 & 25.01 &  &  & 204.75 & 1.8 & 0.08 & b &  & N  $\Delta\mu$\\
HD166553 & 20.25 & 1.13 &  &  & 7.55 & 1.22 & 0.72 & b &  & RV (*) \\
HD167215 & 12.2 &  & 0.37 & 1.56 & 0.37 & 1.15 & 0.09 & c & G  $\Delta\mu$ (*) & SB $\Delta\mu$ \\
HD167389 & 28.8 & 94.18 &  &  & 498.32 & 1.01 & 0.33 & b &  & N \\
HD167665 & 32.01 &  & 0.34 & 5.47 & 1.41 & 1.11 & 0.05 & c & SB & SB  $\Delta\mu$(*) \\
HD169586 & 20.93 &  & 0.35 & 4.3 & 0.83 & 1.29 & 0.68 & c & $\Delta\mu$ & SB  $\Delta\mu$(*) \\
HD169822 & 29.98 &  & 0.48 & 0.84 & 0.13 & 0.91 & 0.3 & c & SB (*) &  \\
HD170778 & 27.14 & 20.19 &  &  & 106.64 & 0.88 & 0.4 & b &  & N  $\Delta\mu$\\
HD170469 & 16.59 & 43.19 &  &  & 406.22 & 1.6 & 0.45 & b &  & N \\
HD174457 & 17.35 &  & 0.23 & 1.9 & 0.6 & 1.07 & 0.06 & c & SB & $\Delta\mu$(*) \\
HD175345 & 24.2 &  & 0.75 & 0.92 & 0.06 & 1.17 & 0.48 & c &  & SB  $\Delta\mu$(*) \\
HD176982 & 9.78 & 15.46 &  &  & 248.72 & 1.43 & 0.38 & b &  & N \\
HD179140 & 18.17 & 0.5 &  &  & 3.53 & 1.12 & 0.82 & c &  & \\
HD179957 & 39.6 & 7.2 &  &  & 21.29 & 1.01 & 1.03 & a &  & $\Delta\mu$ \\

HD179958 & 39.6 & 7.2 &  &  & 21.53 & 1.03 & 1.01 & a &  & $\Delta\mu$ \\
HD180684 & 17.85 &  &  & 52.28 & 7.94 & 1.5 & 0.5 & d &  & N  $\Delta\mu$ \\
HD181655 & 39.65 & 34.6 &  &  & 140.61 & 2.02 & 0.45 & b &  & N \\

HD181234 & 20.87 & 5.17 &  &  & 39.5 & 1.16 & 0.28 & b &  & N \\
HD182488 & 64.06 &  & 0.57 & 21.1 & 3.16 & 1.28 & 0.04 & c &  & N  $\Delta\mu$\\
HD184385 & 47.85 & 37.06 &  &  & 134.25 & 0.88 & 0.09 & b &  & N \\
HD184860 & 35.36 &  & 0.67 & 1.4 & 0.15 & 0.77 & 0.03 & c & SB (*) & $\Delta\mu$\\
HD185395 & 54.23 & 116.71 &  &  & 343.67 & 1.34 & 0.32 & b &  & $\Delta\mu$\\
HD187691 & 51.19 & 21.49 &  &  & 66.42 & 1.37 & 0.35 & b & (*) & \\
HD190067 & 51.83 & 2.86 &  &  & 9.58 & 0.8 & 0.08 & b & (*) & $\Delta\mu$\\
HD190360 & 62.44 & 178.02 &  &  & 465.97 & 1.01 & 0.2 & b & (*) &  \\
HD190406 & 56.43 & 0.8 & 0.5 & 18.3 & 3.25 & 1.09 & 0.07 & c & RV & $\Delta\mu$ (*) \\
HD190771 & 52.59 &  & 0.53 & 8.61 & 1.39 & 1.07 & 0.1 & e & RV  $\Delta\mu$ & $\Delta\mu$ \\
HD191408 & 166.23 & 7.1 &  &  & 6.44 & 0.69 & 0.24 & b &  & RV  $\Delta\mu$(*) \\
HD191785 & 48.85 & 103.84 &  &  & 326.17 & 0.83 & 0.26 & b &  & \\
HD192020 & 40.87 & 15.16 &  &  & 52.12 & 0.78 & 0.39 & b &  & N \\
HD192343 & 15.85 & 43.27 &  &  & 322.18 & 1.28 & 1.27 & a &  & \\
HD192344 & 15.85 & 43.27 &  &  & 320.72 & 1.27 & 1.28 & a &  & \\
HD193017 & 25.02 & 53.62 &  &  & 364.74 & 0.84 & 0.11 & b &  & N \\
HD193307 & 32.05 & 21.26 &  &  & 102.64 & 1.38 & 0.41 & b &  & N \\
HD194766 & 21.7 & 59.88 &  &  & 257.39 & 1.1 & 2.33 & c &  & \\
HD196390 & 29.67 & 235.08 &  &  & 1283.82 & 1.08 & 0.23 & b &  & N \\
HD195564 & 40.38 & 2.9 &  &  & 10.05 & 1.12 & 0.57 & b &  & $\Delta\mu$\\
HD196068 & 24.98 & 16.62 &  &  & 84.27 & 1.69 & 1.3 & c &  & \\
HD196201 & 24.32 & 0.52 &  &  & 2.7 & 0.87 & 0.68 & b &  & RV  $\Delta\mu$ \\
HD197076 & 47.85 & 125.0 &  &  & 351.33 & 0.99 & 0.6 & c & (*) & \\
HD198089 & 24.77 & 2.62 &  &  & 15.4 & 1.0 & 0.42 & a &  & N \\
HD198387 & 24.05 &  & 0.53 & 6.7 & 1.05 & 1.32 & 0.19 & e & G  $\Delta\mu$ (*) & RV  $\Delta\mu$ \\
HD198802 & 22.62 &  &  & 107.84 & 19.79 & 1.6 & 0.03 & d &  & N  $\Delta\mu$ \\
HD199598 & 31.18 &  & 0.0 & 15.1 & 2.63 & 1.15 & 0.11 & c & $\Delta\mu$ & SB  $\Delta\mu$ (*) \\
HD200565 & 15.29 &  & 0.5 & 9.8 & 1.6 & 1.06 & 0.21 & c & RV  $\Delta\mu$ & SB  $\Delta\mu$ \\  
HD200538 & 18.24 & 63.63 &  &  & 615.79 & 1.34 & 0.1 & b &  & N  $\Delta\mu$\\
HD201203 & 11.59 &  &  & 27.86 & 4.52 & 2.23 & 0.47 & d &  & N  $\Delta\mu$ \\  
HD206387 & 19.73 & 2.96 &  &  & 18.55 & 1.2 & 1.01 & b &  & \\
HD206860 & 55.16 & 43.2 &  &  & 143.71 & 1.07 & 0.02 & c &  & \\
HD207700 & 26.02 & 362.17 &  &  & 2244.49 & 1.4 & 0.31 & b &  & N RV $\Delta\mu$(*) \\
HD208776 & 26.07 &  & 0.27 & 4.2 & 0.97 & 1.14 & 0.51 & c & SB & $\Delta\mu$ \\ 
HD208998 & 27.48 & 21.65 &  &  & 132.57 & 1.34 & 0.2 & b &  & N  $\Delta\mu$\\\bottomrule
\end{tabular}}  \end{table}

\begin{table}[H]\ContinuedFloat
\centering \small
\caption{{\em Cont.}} 
\label{tab:binary_UD}
\scalebox{0.85}[0.85]{
\begin{tabular}{lccccccclll}
\toprule
\textbf{ID}          &   \textbf{plx  (mas)}&    \textbf{ rho (\boldmath{$\prime\prime$}) }	&  \textbf{ecc}      &   \textbf{a (au)}     & \textbf{a\boldmath{$_{crit}$} (au)} & \textbf{M\boldmath{$_A$}  M\boldmath{$_{\odot}$} }     & \textbf{M\boldmath{$_B$}   M\boldmath{$_{\odot}$}}     & \textbf{M\boldmath{$_{Flag}$}} & \textbf{BD07 }     & \textbf{Notes} \\
\midrule
HD211681 & 13.82 &  & 0.0 & 5.3 & 0.94 & 1.31 & 0.08 & c &  & N  SB  $\Delta\mu$\\
HD212168 & 42.72 & 20.1 &  &  & 54.52 & 1.06 & 1.12 & c &  & (*) \\
HD212330 & 48.89 &  & 0.21 & 16.21 & 4.64 & 1.12 & 0.27 & c &  & RV $\Delta\mu$(*) \\
HD212708 & 26.91 & 1.01 &  &  & 6.42 & 1.06 & 0.13 & c &  & N  RV $\Delta\mu$ (*) \\
HD213519 & 24.36 & 62.37 &  &  & 385.16 & 1.05 & 0.37 & b &  & \\
HD214953 & 41.97 & 7.43 &  &  & 24.64 & 1.13 & 0.59 & b &  & RV, $\Delta\mu$(*) \\
HD215578 & 4.21 &  & 0.35 & 17.78 & 3.17 & 1.02 & 0.73 & c & RV (*) & SB $\Delta\mu$ (*) \\
HD215648 & 62.4 & 11.05 &  &  & 27.0 & 1.26 & 0.41 & b &  & \\
HD216625 & 23.78 &  &  & 64.44 & 10.69 & 1.39 & 0.24 & d &  & N  $\Delta\mu$\\
HD217004 & 15.62 & 9.17 &  &  & 74.9 & 1.27 & 0.95 & b &  & \\
HD217165 & 22.74 &  & 0.0 & 5.1 & 0.92 & 1.1 & 0.04 & c &  & N  SB  $\Delta\mu$ \\
HD217958 & 16.15 & 1.25 &  &  & 12.15 & 1.15 & 0.31 & c &  & N  $\Delta\mu$(*) \\
HD218101 & 25.84 &  & 0.58 & 6.67 & 0.88 & 1.26 & 0.23 & e & $\Delta\mu$ & SB  $\Delta\mu$ \\
HD218235 & 23.82 & 17.9 &  &  & 135.3 & 2.41 & 0.11 & b &  & N \\
HD218261 & 33.26 &  &  & 80.23 & 12.54 & 1.15 & 0.32 & d &  & N  $\Delta\mu$ \\
HD218730 & 27.24 & 41.85 &  &  & 251.13 & 0.91 & 0.18 & b &  & N \\
HD218868 & 42.45 & 50.4 &  &  & 197.04 & 0.93 & 0.16 & b &  & N  $\Delta\mu$\\
HD219542A & 18.36 & 5.37 &  &  & 34.72 & 1.08 & 1.05 & c & (*) &  \\
HD219542B & 18.19 & 5.37 &  &  & 34.48 & 1.05 & 1.08 & c & (*) &  \\
HD219834 & 44.9 & 12.33 &  &  & 24.22 & 0.74 & 1.87 & c & (*) &  \\ 
HD220077 & 12.87 & 0.66 &  &  & 6.09 & 1.09 & 1.06 & b &  & \\
HD221146 & 27.6 &  &  & 82.34 & 14.56 & 1.66 & 0.12 & d &  & N  $\Delta\mu$ \\
HD221830 & 29.9 & 8.13 &  &  & 39.58 & 0.95 & 0.4 & b &  & \\
HD223084 & 26.95 & 0.16 &  &  & 0.8 & 1.09 & 0.67 & c & RV  $\Delta\mu$ & \\
HD223238 & 22.32 & 24.92 &  &  & 196.70 & 1.29 & 0.1 & b &  & N \\
HD223691 & 13.98 & 13.6 &  &  & 146.97 & 1.6 & 0.55 & c &  & N (*) \\
\bottomrule
\end{tabular}  }
\\
\begin{tabular}{@{}c@{}} 
\multicolumn{1}{p{\textwidth -.75in}}{\footnotesize { \textbf{Remarks:} 
\textbf{N:} System not listed in FV05;
\textbf{SB:} Spectroscopic Binaries; 
\textbf{RV:} Stars with RV linear trends (see~\citep[][]{Nidever2002}); 
\textbf{G:} Stars with accelerating proper motions in Hipparcos. (see \citep{2005AJ....129.2420M}).
\textbf{$\Delta\mu$:} Stars with discrepant proper motion between Hipparcos and Thyco II (see \citep{2005AJ....129.2420M}), or between Gaia DR2 and Thyco II and/or TGAS, see Section~\ref{sec:newdmu} for details)}. 
 \textbf{References}: 
\textbf{HD 4614:} \citet{1983PUSNO..24g...1W}; 
\textbf{HD 4747:} \mbox{\citet{peretti2019}}; 
\textbf{HD 5470:} \mbox{\citet{Patel2007}}; 
\textbf{HD 8673:} \mbox{\citet{roberts2015}}; 
\textbf{HD 8765:} \citet{Patel2007}; 
\textbf{HD 10360/61} \mbox{\citet{1983PUSNO..24g...1W}}; 
\textbf{HD 11112:} \citet{rodigas2016}; 
\textbf{HD 12414:} \citet{gentilefusillo2019}; 
\textbf{HD 13445:} \citet{Lagrange2006,DesideraBarbieri2007}; 
\textbf{HD 13507:} \mbox{\citet{Perrier2003}}; 
\textbf{HD 13531:} \mbox{\citet{metchev}}; 
\textbf{HD 13612B:} \mbox{\citet{1967AJ.....72..899W, DM91}};
\textbf{HD 16160:} \citet{Allen2000}, \mbox{\citet{2000AJ....120.2082G}}; 
\textbf{HD 16895:} \mbox{\citet{1983PUSNO..24g...1W}}; 
\textbf{HD 18445:} \citet{DM91,Halbwachs2000,2001ApJ...562..549Z};
\textbf{HD 18907:} \mbox{\citet{Jenkins2015}} 
\textbf{HD 19467:} \mbox{\citet{crepp2014}} 
\textbf{HD 20782:} \citet{DesideraBarbieri2007}; 
\textbf{HD 23439:} \mbox{\citet{Allen2000}}; 
\textbf{HD 26491:} \mbox{\citet{Jenkins2015}}; 
\textbf{HD 27442:} \citet{Chauvin2006,DesideraBarbieri2007};  
\textbf{HD 29461:} \mbox{\citet{Patel2007}}; 
\textbf{HD 29836:} \citet{1981AJ.....86..588G};
\textbf{HD 30339:} \mbox{\citet{Nidever2002}}; 
\textbf{HD 30649:} \citet{Nidever2002,2002yCat.1274....0D}; 
\textbf{HD 31412:} \citet{Eggenberger2007,Patel2007}; 
\textbf{HD 33636:} \mbox{\citet{bean2007}}; 
\textbf{HD 35956:} \mbox{\citet{2002ApJ...568..352V}};
\textbf{HD 38529:} \citet{Reffert2006, DesideraBarbieri2007};
\textbf{HD 39213:} \mbox{\citet{Jenkins2015}}; 
\textbf{HD 39587:} \mbox{\citet{Nidever2002}}; 
\textbf{HD 40979:} \mbox{\citet{Mugrauer2007}} 
\textbf{HD 42024:} \mbox{\citet{Jenkins2015}}; 
\textbf{HD 43587:} \citet{2002ApJ...568..352V,DM91,2003ApJ...582.1011S}; 
\textbf{HD 43834:} \mbox{\citet{Eggenberger2007}}; 
\textbf{HD 44120:} \mbox{\citet{gentilefusillo2019}}; 
\textbf{HD 45701:} \mbox{\citet{Kane2019}}; 
\textbf{HD 51929:} \mbox{\citet{Kane2019}}; 
\textbf{HD 64184:} \mbox{\citet{Jenkins2015}}; 
\textbf{HD 64468:} \citet{2002ApJ...568..352V};
\textbf{HD 65430:} \mbox{\citet{Nidever2002}}; 
\textbf{HD 65907:} \mbox{\citet{tokovinin2014}};
\textbf{HD 72760:} \mbox{\citet{metchev}};  
\textbf{HD 72780:} \mbox{\citet{Patel2007}}; 
\textbf{HD 73668:} \mbox{\citet{Patel2007}}; 
\textbf{HD 74014:} \mbox{\citet{sahlmann2011}}; 
\textbf{HD 77407:} \mbox{\citet{Mugrauer2004}}, \mbox{\citet{metchev}}; 
\textbf{HD 88218:} \mbox{\citet{tokovinin2014}}; 
\textbf{HIP 52940:} \mbox{\citet{Nidever2002}}; 
\textbf{HD 97334:} \mbox{\citet{2005AJ....129.2849B}};
\textbf{HD 100623:} \mbox{\citet{gentilefusillo2019}}; 
\textbf{HD 101177:} \citet{DM91,2003ApJ...582.1011S};
\textbf{HD 105113:} \mbox{\citet{tokovinin2014}}; 
\textbf{HD 107148:} \mbox{\citet{gentilefusillo2019}}; 
\textbf{HD 114174:} \mbox{\citet{bacchus2017}}; 
\textbf{HD 120136:} \mbox{\citet{justesen2019}};  
\textbf{HD 120690:} \citet{Jenkins2015,dossantos2017}; 
\textbf{HD 121384:} \mbox{\citet{Jenkins2015}}; 
\textbf{HD 122742:} \mbox{\citet{Nidever2002}}; 
\textbf{HD 126614:} \mbox{\citet{howard2010}}; 
\textbf{HD 128620 - HD 128627:} \mbox{\citet{1983PUSNO..24g...1W}}; 
\textbf{HD 130948:} \mbox{\citet{Dupuy2009}}; 
\textbf{HD 131156:} \mbox{\citet{1983PUSNO..24g...1W,DM91}}; 
\textbf{HD 131511:} \mbox{\citet{Nidever2002}}; 
\textbf{HD 131923:} \mbox{\citet{Jenkins2015}}; 
\textbf{HD 131977:} \mbox{\citet{torres2010}}; 
\textbf{HD 135101:} \mbox{\citet{Desidera2004}};
\textbf{HD 136118:} \mbox{\citet{martioli2010}}; 
\textbf{HD 137510:} \mbox{\citet{diaz2012}}; 
\textbf{HD 137778:} \mbox{\citet{tokovinin2016}}; 
\textbf{HD 139323:} this paper; 
\textbf{HD 140901:} \mbox{\citet{gentilefusillo2019}}; 
\textbf{HD 140913:} \mbox{\citet{Nidever2002}};
\textbf{HD 142229:} \mbox{\citet{Patel2007}}; 
\textbf{HD 145825:} \mbox{\citet{Kane2019}}
\textbf{HD 146362 B:} \mbox{\citet{raghavan2009}};
\textbf{HD 150248:} \mbox{\citet{Jenkins2015}}; 
\textbf{HD 150554:} \mbox{\citet{Patel2007}}; \mbox{\citet{metchev}};     
\textbf{HD 156274B:} \mbox{\citet{Jenkins2015}}; 
\textbf{HD 158733:} \mbox{\citet{Jenkins2015}}; 
\textbf{HD 161797:} \citet{1983PUSNO..24g...1W,Nidever2002,2006AJ....132..177W,2005AJ....129.2420M}; 
\textbf{HD 162255:} \mbox{\citet{Jenkins2015}}; 
\textbf{HD 167215:} \mbox{\citet{diaz2012}};
\textbf{HD 167665:} \mbox{\citet{Patel2007}}; 
\textbf{HD 169586:} \mbox{\citet{Jenkins2015}}; 
\textbf{HD 169822:} \mbox{\citet{2002ApJ...568..352V}};
\textbf{HD 174457:} \mbox{\citet{Nidever2002}};
\textbf{HD 175345:} \mbox{\citet{Jenkins2015}}; 
\textbf{HD 177830:} \mbox{\citet{Eggenberger2007}};  
\textbf{HD 178911 B:} \mbox{\citet{Tokovinin2000}}; 
\textbf{HD 179140:} \mbox{\citet{tokovinin2014}}; 
\textbf{HD 182488:} \mbox{\citet{bowler2018}}; 
\textbf{HD 184860:} \mbox{\citet{2002ApJ...568..352V}};
\textbf{HD 190406:} \mbox{\citet{crepp2012}};
\textbf{HD 194766:} \mbox{\citet{tokovinin2014}}; 
\textbf{HD 196050:} \mbox{\citet{Eggenberger2007}}; 
\textbf{HD 196068:} \mbox{\citet{marmier2013}}; 
\textbf{HD 197076:} \mbox{\citet{DM91}};
\textbf{HD 196885:} \mbox{\citet{chauvin2011}};  
\textbf{HD 199598:} \mbox{\citet{Patel2007}}; 
\textbf{HD 200565:} \mbox{\citet{dossantos2017}}; 
\textbf{HD 206860:} \mbox{\citet{luhman2007}}
\textbf{HD 208776:} \mbox{\citet{Nidever2002}}; 
\textbf{HD 211681:} \mbox{\citet{Patel2007}}; 
\textbf{HD 212168:} \mbox{\citet{tokovinin2014}}; 
\textbf{HD 212330:} \mbox{\citet{Kane2019}}; 
\textbf{HD 212708:} \mbox{\citet{Kane2019}}; 
\textbf{HD 215578:} \mbox{\citet{Patel2007}}; 
\textbf{HD 217165:} \mbox{\citet{Patel2007}}; 
\textbf{HD 217958:} \mbox{\citet{Kane2019}}; this paper; 
\textbf{HD 219542:} \mbox{\citet{Desidera2004};} 
\textbf{HD 219834:} \mbox{\citet{docobo2018}};
\textbf{HD 223084:} \mbox{\citet{tokovinin2012}}; 
\textbf{HD 223691:} This paper.
}
\end{tabular}

\end{table}


\section{Notes on Individual Systems}
\label{app:remarks}

{\bf HD 8673:} This star hosts a massive planet in a highly eccentric orbit with a period slightly longer than the UD boundaries \citep{hartmann2010}. Adaptive optics (AO) observations by \cite{roberts2015} revealed a stellar companion at 0.3$^{\prime\prime}$. They estimated a mass of 0.33--0.45 $M_{\odot}$. The orbit is constrained to be a = 35--60 au and e < 0.5. The star also has a $\Delta \mu$ signature, which may be due to the planet and/or the stellar companion, while there are no indications of long-term trends from RV measurements that yielded  planet discovery.

{\bf HD 8765:} Star already considered as binary in BD07 on the basis of the astrometric trend reported by \cite{MakarovKaplan2005}.
The presence of a stellar companion is confirmed by \cite{Patel2007}.

{\bf HD 11112:} Star with an RV trend and $\Delta \mu$. The responsible for the trend was identified by \citep{rodigas2016} as a white dwarf at 2.2” with a probable mass of 0.9--1.1 $M_{\odot}$.

{\bf HD 12414:} New companion identified through Gaia DR2. The blue BP-RP color indicates that it is white dwarf companion. This is further confirmed by the analysis of
\citet{gentilefusillo2019}. Therefore, this star is a new Sirius-like system.

{\bf HD 18907:} Spectroscopic binary (SB) from \cite{Jenkins2015} (very preliminary orbital solution due to the long period). The star has also a $\Delta \mu$ signature.

{\bf HD 19467:} The star has a long-term RV trend, astrometric 
$\Delta \mu$, and direct detection with AO. The~companion is a T-type brown dwarf \citep{crepp2014}.

{\bf HD 26491:} SB from \cite{Jenkins2015} (incomplete orbital solution due to long period). The star was already in BD07 as $\Delta \mu$ binary.

{\bf HD 29461, HD 72780, HD 142229, HD 167665, HD 215578:} Already included in the UD binary sub-sample, on the basis of the RV trend reported by \cite{Nidever2002}, are now confirmed by the work of P07.

{\bf HD 31412:} E07 confirmed that this system is made by a close pair (HD 31412Ab $\rho$ = 0.194$^{\prime\prime}$) orbited by a distant companion, HD 31412B, at 22$^{\prime\prime}$.  HD 31412Aa is likely an early M-dwarf with a mass of 0.42 $M_{\odot}$. This mass estimate is in agreement with P07 which reported  a mass between 0.36--0.41 $M_{\odot}$ for HD 31412Aa.

{\bf HD 33636:} Observations from the Hubble Space Telescope (HST) indicate that the RV planet (msini=9 $M_{Jup}$) is instead a low-mass star (see \citep[][]{bean2007} for details).

{\bf HD 39213:} SB with an orbital solution from \cite{Jenkins2015}. This also has a $\Delta \mu$ signature. The minimum mass from the RV orbital solution is formally in the brown dwarf (BD) regime (69.7 $M_{Jup}$).

{\bf HD 40979:} The comoving companion reported by \cite{Halbwachs1986} was found to be a close pair ($\rho$ = 3.9$^{\prime\prime}$) by~\cite{Mugrauer2007}, with a total mass of 1.21 $M_{\odot}$ (0.833 $M_{\odot}$ for HD 40979B and 0.380 $M_{\odot}$ for HD 40979C).

{\bf HD 42024:} Triple system. SB with an orbital solution from \cite{Jenkins2015}. 
The companion has minimum mass in the substellar regime (68.9 $M_{Jup}$). There is also a visual companion at 2.67$^{\prime\prime}$ from Gaia.

{\bf HD 44120:} The companion is a white dwarf. The F-type star HD 44105 at 34$^{\prime\prime}$ is not physically associated, according to Gaia DR2 astrometric parameters.

{\bf HD 45701:} Included in BD07 as $\Delta \mu$ binary.
Full orbital solution from RV is presented in \cite{Kane2019}, with~mass
in the stellar regime.  

\textls[-5]{{\bf HD 50281:} The two components of this binary system
show a large difference in proper motion ($\Delta \mu_{\alpha}=32.6$ mas/yr; $\Delta \mu_{\delta}=9.8$ mas/yr). Nevertheless, we consider
it as a binary system, considering the very similar parallax in Gaia
DR2 (114.30 mas for the primary and 114.41 mas for the secondary),
the~large proper motion (--543.62 and --3.49 mas/yr for the primary),
and the very similar radial velocities.
The origin of the large proper motion difference is unclear, as the pair has  a projected separation of 510 au, ruling out significant
orbital motion, and there are no indications of close companions around
either of the components (HD50281B also has high-precision RV monitoring
from \cite{Butler2017}). }

{\bf HD 51929:} Included in BD07 as $\Delta \mu$ binary. The 
companion likely responsible for the astrometric acceleration was 
detected  using both RV (long-term trend) and imaging by \cite{Kane2019}.

{\bf HD 52447:} Star with a long-term RV trend \citep{Kane2019}
and Gaia2/TGAS $\Delta \mu$.
From the properties of the long-term trend, it results that, most likely, the 
companion is stellar.


{\bf HD 65277:} The star has a close source (HD 65277B) at 5$^{\prime\prime}$, which we consider as
a physical companion, as the $\mu_{\alpha}$ and parallax are very similar. The large discrepancy in $\mu_{\delta}$ (23 mas/yr) could be due to the presence of an additional companion and/or to the orbital motion of the pair (projected separation of about 90 au). The secondary has a highly significant astrometric excess noise from Gaia DR2, suggesting that it could have an additional companion. 
The UD target (the primary) has a shallow RV trend, which is compatible with the presence of the secondary.

{\bf HD 73668:} Previously included in BD07 because of a companion reported  at 26.50$^{\prime\prime}$, it was found by \cite{Patel2007} to have an additional companion with a period of 20.4 years and a mass of 164 $M_{Jup}$.

{\bf HD 80913:} Star with a long-term RV trend \citep{Kane2019} and Gaia2/TGAS $\Delta\mu$.
From the properties of the long-term trend, it results that, most likely, the 
companion has mass larger than 20 $M_{Jup}$.

{\bf HIP 52940} and {\bf HIP 52942:} Probable
quadruple system. The two components have a separation of 17.5$^{\prime\prime}$.
HIP 52940 is a close visual and spectroscopic binary.
At a wide separation (437$^{\prime\prime}$), there is the white dwarf 
SDSS J104937.10+124827.8, which has similar values of parallax
with some discrepancy in proper motion.

{\bf HD 100623:} Included in BD07 as a wide binary (separation of 17$^{\prime\prime}$), confirmed by Gaia DR2. The~companion is a white dwarf. The star has also a long-term RV trend \citep{Kane2019} and $\Delta \mu$ signature, which might be due to the WD (projected separation of about 160 au).

{\bf HD 105113:} The secondary HD 105113B is 
a close binary, identified as SB2 by \cite{tokovinin2014}
and close visual binary by \cite{mason2018}.

{\bf HD 107148:} New companion identified through
Gaia DR2. The blue BP-RP color indicates that it is white dwarf
companion. This is further confirmed by the analysis of
\citet{gentilefusillo2019}. Therefore, this star is 
a new Sirius-like system. The star also has a planet
below the RV amplitude threshold of the UD sample.

{\bf HD 114174:} White dwarf companion at 0.7$^{\prime\prime}$ detected by \cite{crepp2013}. It is responsible for the RV trend and astrometric signature.
Detailed studies show some ambiguities in the properties of the white dwarf \citep{matthews2014,bacchus2017}.
We adopt a mass of 1.20 and the corresponding orbital solution from \cite{bacchus2017}.  

{\bf HD 117939:} New companion from Gaia DR2 at 3.29$^{\prime\prime}$ = 99.4 au.
This object could be responsible for the long-term RV trend found by  \cite{Kane2019}.

{\bf HD 120690:} SB with an orbital solution from \cite{Jenkins2015,dossantos2017}. 
The companion also has direct detection, which yields a mass close to the minimum mass from the spectroscopic orbit \citep{tokovinin2014}. It was already included in BD07 as a $\Delta\mu$ binary.

\textls[-15]{{\bf HD 121384:} SB with an orbital solution from \cite{Jenkins2015}. 
There is also a wide companion at 34.2$^{\prime\prime}$ from Gaia}. 

{\bf HD 126614:} Companion at 0.49$^{\prime\prime}$ detected by \cite{howard2010}. It is  responsible for the RV trend and astrometric signature. There is an additional comoving object at 41.85 from Gaia.

{\bf HD 131923:} SB from \cite{Jenkins2015} (incomplete orbital solution due to the long period). The star was already in BD07 as a $\Delta \mu$ binary and a star with Hipparcos astrometric acceleration.

{\bf HD 131977:} The 28$^{\prime\prime}$ companion to the UD target is itself a close binary with individual masses adopted from \cite{torres2010}. There is also an additional companion, the brown dwarf GJ 570D, at 262$^{\prime\prime}$.

{\bf HD 137778:} 
The large difference in parallax and proper motion between the components is likely due to the binarity of the companion, the G star HD 137763, which is a close visual and spectroscopic binary with an orbital solution \citep{tokovinin2016}.

{\bf HD 139323:} The secondary (HD 139341) is a close visual and spectroscopic binary.

{\bf HD 140901:} The companion is a WD \citep{gentilefusillo2019}.

{\bf HD 145825:} New binary from the RV orbital solution
\citep{Kane2019}, with mass
in the stellar regime.  

{\bf HD 150248:} SB with orbital solution from \cite{Jenkins2015}. The star was already in BD07 as a $\Delta\mu$ binary.

{\bf HD 150554:} BD07 reported the companion listed in \cite{Dupuy2009} at 11.60$^{\prime\prime}$, but P07 discovered a further low-mass companion (Msini $\sim 68.8 M_{Jup}$) with a period of 11.8 years.

{\bf HD 156274B:} SB from \cite{Jenkins2015} (incomplete orbital solution due to the long period). The star is part of a triple system with HD 156274A at 8.65$^{\prime\prime}$,

{\bf HD 158783:} SB with an orbital solution from \cite{Jenkins2015}. The star also has a $\Delta \mu$ signature. 

{\bf HD 162255:} SB with an orbital solution from \cite{Jenkins2015}. The star also has a $\Delta \mu$ signature. There is also a wide companion at 85$^{\prime\prime}$, which is likely a white dwarf from  the moderately blue BP-RP color and lack of detection in 2MASS.

{\bf HD 166553:} Already in the UD sample as a close visual binary (separation of 1.4$^{\prime\prime}$). Confirmed from the RV long-term trend and direct imaging detection  \citep{Kane2019}.

{\bf HD 169586:} SB with an orbital solution from \cite{Jenkins2015}. The star was already included in BD07 as a $\Delta \mu$~binary. 

{\bf HD 175345:} SB with an orbital solution from \cite{Jenkins2015}. This also has a $\Delta \mu$ signature from Gaia. There is also a companion
at a wider separation (6.19$^{\prime\prime}$).

{\bf HD 191408:} Included in BD07 as a visual binary (separation of $\rho = 7.1^{\prime\prime}$).
The companion (GJ 783B; spectral type M3.5) is not included in Gaia DR2, most likely because of the brightness of the primary (mag 5) and very large proper motion.
The star also an RV long-term trend \citep{Kane2019} and  significant $\Delta \mu$.
The known companion (projected separation of 42 au) is likely responsible for
these dynamical signatures.

{\bf HD 196885:} The occurrence of a planetary companion fulfilling the UD definition was confirmed by  \cite{Correia2008,fischer2009}. The authors of \cite{chauvin2011} derived a binary orbital
solution including the AO observations and the RV trend.

{\bf HD 196050:} The comoving companion reported in BD07 was found by \cite{Eggenberger2007} to be a close pair made of an M1.5--M4.5 dwarf with a mass of $0.29\pm 0.02 M_{\odot}$ plus an M2.5--M5.5 dwarf of $0.19\pm 0.02 M_{\odot}$.

{\bf HD 199598:} Already considered as a binary in BD07 on the basis of the astrometric trend reported by \cite{MakarovKaplan2005}, and is now confirmed by the work of \cite{Patel2007}.

{\bf HD 207700:} New very wide visual binary from Gaia DR2 (projected separation of 362$^{\prime\prime}$). The~star also has a long-term RV trend from 
\cite{Kane2019} and proper motion difference from GaiaDR2-TGAS.
This close companion may be of planetary mass, and is therefore not included in
this paper as a binary companion. 

{\bf HD 212168:} The wide companion at 20.1$^{\prime\prime}$ from the UD target is itself a close (projected separation of 0.077$^{\prime\prime}$) binary \citep{tokovinin2014}.  

\textls[-15]{{\bf HD 212330:} Included in BD07 as a wide binary because
of the presence of a wide companion at 81$^{\prime\prime}$. The physical association is
confirmed by Gaia DR2. An additional companion at closer separation was 
identified by \cite{Kane2019} with an RV orbital solution and detection by direct imaging. 
The system is then triple.}

{\bf HD 212708:} New binary from RV long-term trend and direct imaging detection  \citep{Kane2019}.

{\bf HD 214953:} Originally in BD07 as a visual binary (separation of 7$^{\prime\prime}$)
It also has an RV long-term trend \citep{Kane2019} and a $\Delta \mu$ signature.
These trends might be due to the known companion or to an additional one of
much smaller mass at closer separation.

{\bf HD 217958:} New binary from the RV long-term trend and direct imaging detection \citep{Kane2019}, with mass in the stellar regime. 

{\bf HD 223691:} The new companion detected by Gaia at 13" results in a probable white dwarf from
the position on the color magnitude diagram ($M_G=15.11$, BP-RP=1.33), well within the locus of white dwarfs from \citet{gentilefusillo2019}.
The modest amount of photometric contamination (phot bprp excess factor = 1.91) supports the classification based on 
BP-RP.  The star then results in a new Sirius-like system. We adopt a mass of 0.55 M$_{\odot}$ for the companion, typical for this kind of star.

\reftitle{References}




\end{document}